\documentclass[nojss]{jss}
\usepackage{rccol} 
\usepackage{booktabs}
\usepackage{amsmath} \usepackage{amssymb} \usepackage{bm}
\author{
Garth Tarr\\University of Newcastle \And Samuel Mueller\\University of Sydney \And Alan H.\ Welsh\\Australian National University
}
\title{\pkg{mplot}: An \proglang{R} Package for Graphical Model Stability and Variable Selection Procedures}

\Abstract{
The \pkg{mplot} package provides an easy to use implementation of model
stability and variable inclusion plots \citep{Mueller:2010, Murray:2013} as well as the adaptive fence \citep{Jiang:2008, Jiang:2009} for linear and generalised linear models. We provide a number of innovations on the standard procedures and address many practical implementation issues including the addition of redundant variables, interactive visualisations and approximating logistic models with linear models. An option is provided that combines our bootstrap approach with \pkg{glmnet} for higher dimensional models.  The plots and graphical user interface leverage state of the art web technologies to facilitate interaction with the results. The speed of implementation comes from the \pkg{leaps} package and cross-platform multicore support.
}

\Plainauthor{Garth Tarr, Samuel Mueller, Alan H. Welsh}
\Plaintitle{mplot: An R Package for Graphical Model Stability and Variable Selection Procedures}
\Shorttitle{\pkg{mplot}: Graphical Model Stability and Variable Selection in \proglang{R}}
\Plainkeywords{model selection, variable selection, linear models, mixed models, generalised linear models, fence, R}

\Address{
    Garth Tarr\\
  University of Newcastle\\
  School of Mathematical and Physical Sciences \newline University of
  Newcastle \newline Callaghan NSW 2308\\
  E-mail: \href{mailto:garth.tarr@newcastle.edu.au}{\nolinkurl{garth.tarr@newcastle.edu.au}}\\
  
      Samuel Mueller\\
  University of Sydney\\
  School of Mathematics and Statistics \newline University of Sydney
  \newline Sydney NSW 2006\\
  E-mail: \href{mailto:samuel.mueller@sydney.edu.au}{\nolinkurl{samuel.mueller@sydney.edu.au}}\\
  
      Alan H Welsh\\
  Australian National University\\
  Mathematical Sciences Institute \newline Australian National University
  \newline Canberra ACT 2601\\
  E-mail: \href{mailto:alan.welsh@anu.edu.au}{\nolinkurl{alan.welsh@anu.edu.au}}\\
  
  }

\begin{document}

\section{Graphical tools for model selection}\label{graphical-tools-for-model-selection}

In this article we introduce the \pkg{mplot} package in \proglang{R}, which provides a suite of interactive visualisations and model summary statistics for researchers to use to better inform the variable selection process \citep{Tarr:2015c,R-Core-Team:2015aa}. The methods we provide rely heavily on various bootstrap techniques to give an indication of the stability of selecting a given model or variable and even though not done here, could be implemented with resampling methods other than the bootstrap, for example cross validation. The `{\bf m}' in \pkg{mplot} stands for model selection/building and we anticipate that in  future more graphs and methods will be added to the package to further aid better and more stable building of regression models.  The intention is to encourage researchers to engage more closely with the model selection process, allowing them to pair their experience and domain specific knowledge with comprehensive summaries of the relative importance of various statistical models.

Two major challenges in model building are the vast number of models to choose from and the myriad of ways to do so.  Standard approaches include stepwise variable selection techniques and more recently the lasso. A common issue with these and other methods is their instability, that is, the tendency for small changes in the data to lead to the selection of different models.

An early and significant contribution to the use of bootstrap model selection is \citet{Shao:1996} who showed that carefully selecting $m$ in an $m$-out-of-$n$ bootstrap drives the theoretical properties of the model selector. \citet{Mueller:2005,Mueller:2009} modified and generalised Shao's $m$-out-of-$n$ bootstrap model selection method to robust settings, first in linear regression and then in generalised linear models. The bootstrap is also used in regression models that are not yet covered by the \pkg{mplot} package, such as mixed models \citep[e.g.,][]{Shang:2008} or partially linear models \citep[e.g.,][]{Mueller:2009b} as well as for the selection of tuning parameters in regularisation methods \citep[e.g.,][]{Park:2014}.

Assume that we have $n$ independent observations $\mathbf{y} = (y_{1},\ldots,y_{n})^{\top}$ and an $n\times p$ full rank design matrix $\mathbf{X}$ whose columns are indexed by $1,\ldots,p$. Let $\alpha$ denote any subset of $p_{\alpha}$ distinct elements from $\{1,\ldots,p\}$. Let $\mathbf{X}_{\alpha}$ be the corresponding $n\times p_{\alpha}$ design matrix and $\mathbf{x}_{\alpha i}^{\top}$ denote the $i$th row of $\mathbf{X}_{\alpha}$.

The \pkg{mplot} package focuses specifically on linear and generalised linear models (GLM). In the context of GLMs, a model $\alpha$ for the relationship between the response $\mathbf{y}$ and the design matrix $\mathbf{X}_{\alpha}$ is specified by
\begin{align}
\mathbb{E}(\mathbf{y}) = h(\mathbf{X}_{\alpha}^{\top}\bm{\beta}_{\alpha}), \text{ and }\operatorname{var}(\mathbf{y}) = \sigma^{2}v(h(\mathbf{X}_{\alpha}^{\top}\bm{\beta}_{\alpha})),
\end{align}
where $\bm{\beta}_{\alpha}$ is an unknown $p_{\alpha}$-vector of regression parameters and $\sigma$ is an unknown scale parameter. Here $\mathbb{E}(\cdot)$ and $\operatorname{var}(\cdot)$  denote the expected value and variance of a random variable, $h$ is the inverse of the usual link function and both $h$ and $v$ are assumed known. When $h$ is the identity and $v(\cdot)=1$, we recover the standard linear model.  

The purpose of model selection is to choose one or more models $\alpha$ from a set of candidate models, which may be the set of all models $\mathcal{A}$ or a reduced model set (obtained, for example, using any initial screening method). Many model selection procedures assess model fit using the generalised information criterion,
\begin{equation}
\textrm{GIC}(\alpha,\lambda) = \hat{Q}(\alpha) + \lambda p_{\alpha}. \label{GIC}
\end{equation}
The $\hat{Q}(\alpha)$ component is a measure of ``description loss'' or ``lack of fit'', a function that describes how well a model fits the data, for example, the residual sum of squares or $-2~\times~\text{log-likelihood}$. The number of independent regression model parameters, $p_{\alpha}$, is a measure of ``model complexity''. The penalty multiplier, $\lambda$, determines the properties of the model selection criterion \citep{Mueller:2013,Mueller:2010}. Special cases, when $\hat{Q}(\alpha)=-2\times\text{log-likelihood}(\alpha)$, include the AIC with $\lambda=2$, BIC with $\lambda=\log(n)$ and more generally the generalised information criterion (GIC) with $\lambda\in\mathbb{R}$ \citep{Konishi:1996}.

The \pkg{mplot} package currently implements ``variable inclusion plots'', ``model stability plots'' and a model selection procedure inspired by the adaptive fence of \citet{Jiang:2008}.  Variable inclusion plots were introduced independently by \citet{Mueller:2010} and \citet{Meinshausen:2010}. The idea is that the best model is selected over a range of values of the penalty multiplier $\lambda$ and the results are visualised on a plot which shows how often each variable is included in the best model.  These types of plots have previously been referred to as stability paths, model selection curves and most recently variable inclusion plots (VIPs) in \citet{Murray:2013}.  An alternative to penalising for the number of variables in a model is to assess the fit of models within each model size. This is the approach taken in our model stability plots where searches are performed over a number of bootstrap replications and the best models for each size are tallied. The rationale is that if there exists a ``correct'' model of a particular model size it will be selected overwhelmingly more often than other models of the same size.  Finally, the adaptive fence was introduced by \citet{Jiang:2008} to select mixed models.  This is the first time code has been made available to implement the adaptive fence and the first time the adaptive fence has been applied to linear and generalised linear models.

This article introduces three data examples that each highlight different aspects of the graphical methods made available by \pkg{mplot}. Sections \ref{sec:ie}-\ref{sec:ig} are based on a motivating example where the true data generating model is known. We use this example to highlight one of the classical failings of stepwise procedures before introducing variable inclusion plots and model stability plots through the \code{vis()} function in Section \ref{sec:vis}. Our implementation of the adaptive fence with the \code{af()} function is presented in Section \ref{sec:af}. 

For all methods, we provide publication quality classical plot methods using base \proglang{R} graphics as well as interactive plots using the \pkg{googleVis} package \citep{Gesmann:2011}. In Section \ref{sec:ig}, we show how to add further utility to these plot methods by packaging the results in a \pkg{shiny} web interface which facilitates a high degree of interactivity \citep{Chang:2015a}.

In Section \ref{sec:timing} we show computing times in a simulation study, varying the number of variables from 5 to 50; we further illustrate the advantage of using multiple core technology.  We then show with two applied examples the practical merit of our graphical tools in Section \ref{sec:examples}.

To conclude, we highlight in Section \ref{sec:conclusion} the key contributions of the three data examples and make some final brief remarks.

\section{Illustrative example}\label{sec:ie}

We will present three examples to help illustrate the methods provided by the \pkg{mplot} package. Two real data sets are presented as case studies in Section \ref{sec:examples}. The first of these is a subset of the diabetes data set used in \citet{Efron:2004} which has 10 explanatory variables and a continuous dependent variable, a measure of disease progression, suitable for use in a linear regression model.  The second is a binomial regression example from \citet{Hosmer:1989book} concerning low birth weight.  

\begin{figure}[t]
\includegraphics[width=\textwidth]{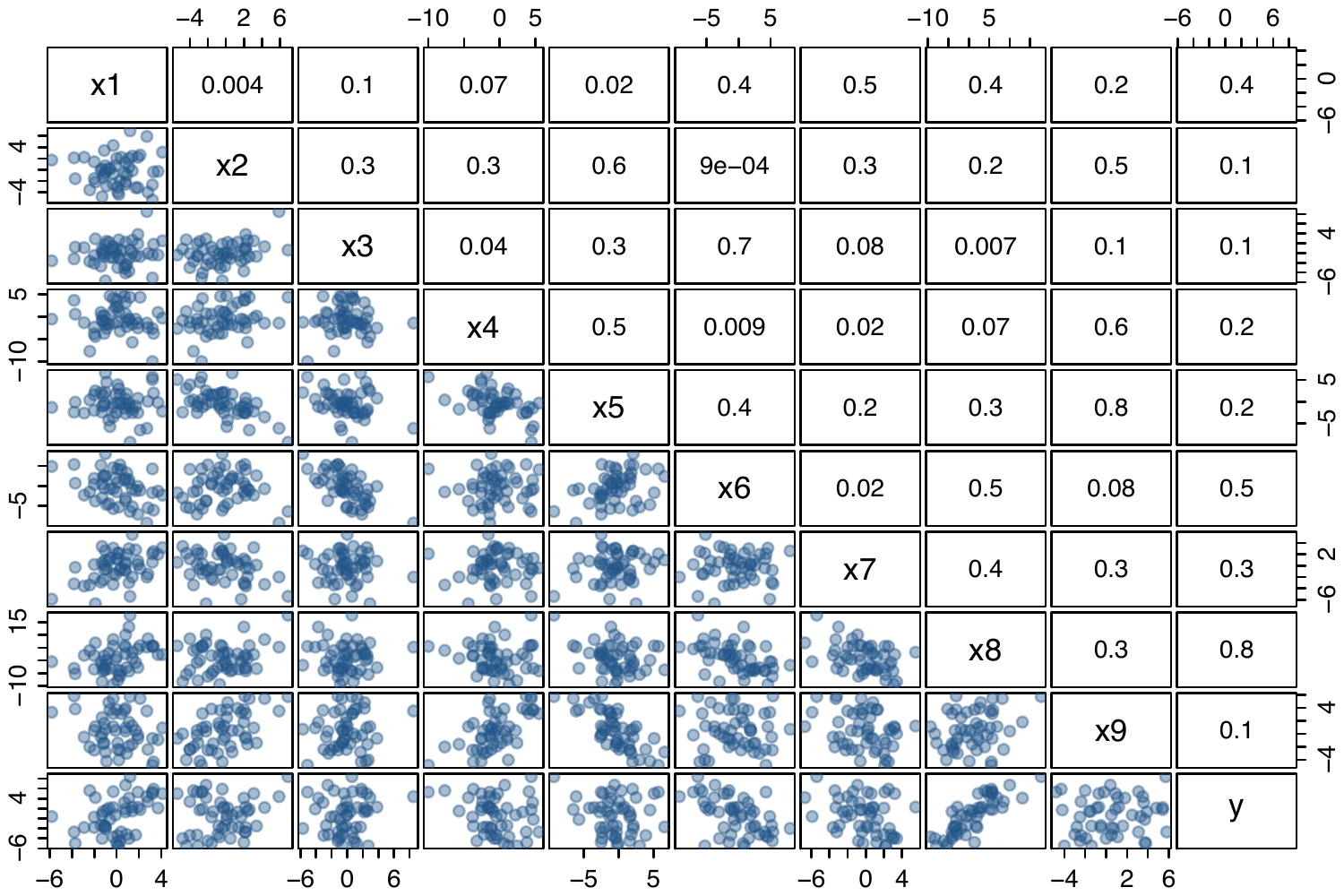}
\caption{Scatterplot matrix of the artificially generated data set with estimated correlations in the upper right triangle.  The true data generating process for the dependent variable is $y=0.6\, x_8 + \varepsilon$ where $\varepsilon\sim\mathcal{N}(0,2^2)$.}
\label{pairsplot}
\end{figure}

The artificially generated data set was originally designed to emphasise statistical deficiencies in stepwise procedures, but here it will be used to highlight the utility of the various procedures and plots provided by  \pkg{mplot}.   The data set and details of how it was generated are provided with the \pkg{mplot} package.
\begin{CodeChunk}
\begin{CodeInput}
R> install.packages("mplot")
R> data("artificialeg", package = "mplot")
R> help("artificialeg", package = "mplot")
\end{CodeInput}
\end{CodeChunk}
A scatterplot matrix of the data and the estimated pairwise correlations are given in Figure~\ref{pairsplot}.  There are no outliers and we have not positioned the observations in a subspace of the artificially generated data set. All variables, while related, originate from a Gaussian distribution.  Fitting the full model yields no individually significant variables.

\begin{CodeChunk}
\begin{CodeInput}
R> require("mplot")
R> data("artificialeg")
R> full.model = lm(y ~ ., data = artificialeg)
R> round(summary(full.model)$coef, 2)
\end{CodeInput}
\begin{CodeOutput}
            Estimate Std. Error t value Pr(>|t|)
(Intercept)    -0.10       0.33   -0.31     0.76
x1              0.64       0.69    0.92     0.36
x2              0.26       0.62    0.42     0.68
x3             -0.51       1.24   -0.41     0.68
x4             -0.30       0.25   -1.18     0.24
x5              0.36       0.60    0.59     0.56
x6             -0.54       0.96   -0.56     0.58
x7             -0.43       0.63   -0.68     0.50
x8              0.15       0.62    0.24     0.81
x9              0.40       0.64    0.63     0.53
\end{CodeOutput}
\end{CodeChunk}

Performing default stepwise variable selection yields a model with all explanatory variables except $x_8$.  As an aside, the dramatic changes in the p-values indicate that there is substantial interdependence between the explanatory variables even though none of the pairwise correlations in Figure~\ref{pairsplot} are particularly extreme.

\begin{CodeChunk}
\begin{CodeInput}
R> step.model = step(full.model, trace = 0)
R> round(summary(step.model)$coef, 2)
\end{CodeInput}
\begin{CodeOutput}
            Estimate Std. Error t value Pr(>|t|)
(Intercept)    -0.11       0.32   -0.36     0.72
x1              0.80       0.19    4.13     0.00
x2              0.40       0.18    2.26     0.03
x3             -0.81       0.19   -4.22     0.00
x4             -0.35       0.12   -2.94     0.01
x5              0.49       0.19    2.55     0.01
x6             -0.77       0.15   -5.19     0.00
x7             -0.58       0.15   -3.94     0.00
x9              0.55       0.19    2.90     0.01
\end{CodeOutput}
\end{CodeChunk}

The true data generating process is, $y = 0.6\,x_{8} + \varepsilon$, where $\varepsilon\sim\mathcal{N}(0,2^2)$.  The bivariate regression of $y$ on $x_{8}$ is the more desirable model, not just because it is the true model representing the data generating process, but it is also more parsimonious with essentially the same residual variance as the larger model chosen by the stepwise procedure.  This example illustrates a key statistical failing of stepwise model selection procedures, in that they only explore a subset of the model space so are inherently susceptible to local minima in the information criterion \citep{Harrell:2001}.  

Perhaps the real problem with of stepwise methods is that they allow researchers to transfer all responsibility for model selection to a computer and not put any real thought into the model selection process.  This is an issue that is also shared, to a certain extent with more recent model selection procedures based on regularisation such as the lasso and least angle regression \citep{Tibshirani:1996,Tibshirani:2004}, where attention focusses only on those models that are identified by the path taken through the model space. In the lasso, as the tuning parameter $\lambda$ is varied from zero to $\infty$, different regression parameters remain non-zero, thus generating a path through the set of possible regression models, starting with the largest ``optimal'' model when $\lambda=0$ to the smallest possible model when $\lambda=\infty$, typically the null model because the intercept is not penalised. The lasso selects that model on the lasso path at a single $\lambda$ value, that minimises one of the many possible criteria (such as 5-fold cross-validation, or the prediction error) or by determining the model on the lasso path that minimises an information criterion (for example BIC).

An alternative to stepwise or regularisation procedures is to perform exhaustive searches of the model space.  While exhaustive searches avoid the issue of local minima, they are computationally expensive, growing exponentially in the number of variables $p$, with more than a thousand models when $p=10$ and a million when $p=20$.  The methods provided in the \pkg{mplot} package and described in the remainder of the article go beyond stepwise procedures by incorporating exhaustive searches where feasible and using resampling techniques to provide an indication of the stability of the selected model.  The \pkg{mplot} package can feasibly handle up to 50 variables in linear regression models and a similar number for logistic regression models when an appropriate transformation (described in Section \ref{sec:bw}) is implemented.  

\section{Model stability and variable inclusion plots}\label{sec:vis}

The main contributions of the \pkg{mplot} package are model stability plots and variable inclusion plots, implemented through the \code{vis()} function, and the simplified adaptive fence for linear and generalised linear models via the \code{af()} function which is discussed in Section \ref{sec:af}.

Our methods generate large amounts of raw data about the fitted models.  While the print and summary output from both functions provide suggestions as to which models appear to be performing best, it is not our intention to have researchers simply read off the ``best'' model from the output.  The primary purpose of these techniques is to help inform a researcher's model selection choice. As such, the real value in using these functions is in the extensive plot methods provided that help visualise the results and get new insights.  This is reflected in the choice of name \code{vis}, short for visualise, as this is the ultimate goal -- to visualise the stability of the model selection process.  

\subsection{Model stability plots}\label{sec:msp}

In order to generate model stability and variable inclusion plots, the first step is to generate a \code{vis} object using the \code{vis()} function.  To generate a \code{vis} object for the artificial data example the fitted full model object along with some optional arguments are passed to the \code{vis()} function.

\begin{CodeChunk}
\begin{CodeInput}
R> lm.art = lm(y ~ ., data = artificialeg)
R> vis.art = vis(lm.art, B = 150, redundant = TRUE, nbest = "all")
\end{CodeInput}
\end{CodeChunk}

\begin{figure}[t]
\centering
\includegraphics[width=0.5\textwidth]{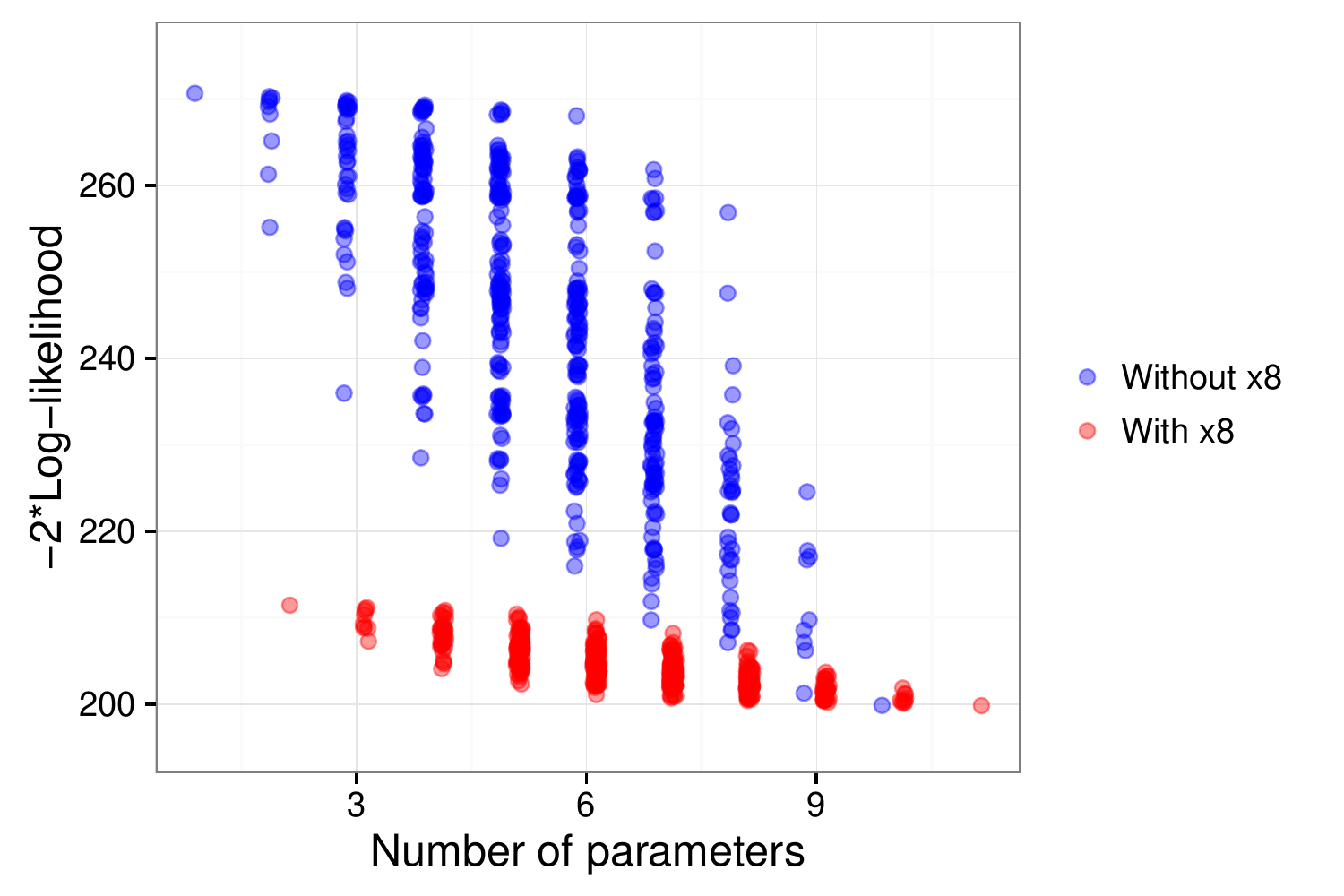}\includegraphics[width=0.5\textwidth]{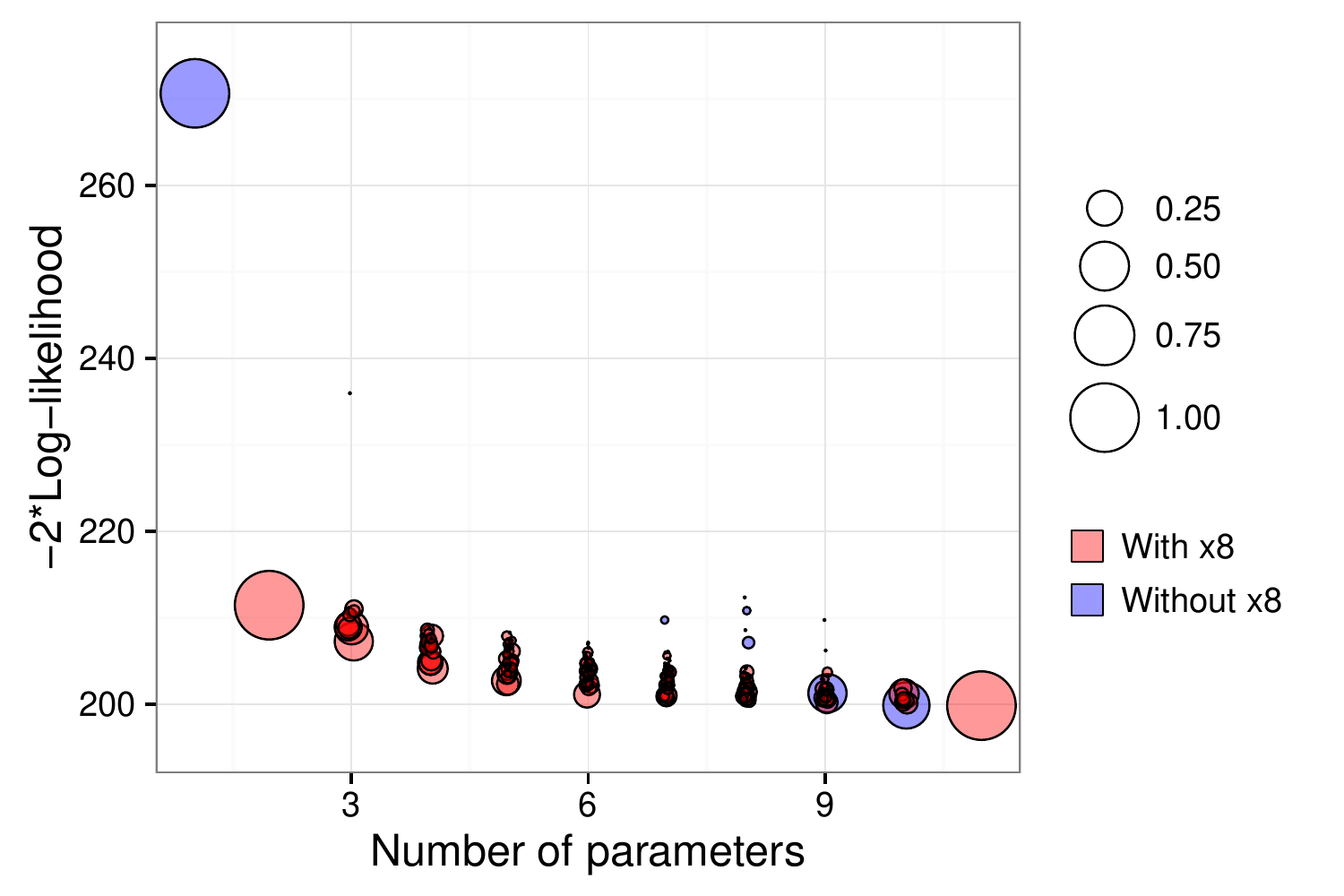}

\includegraphics[width=\textwidth]{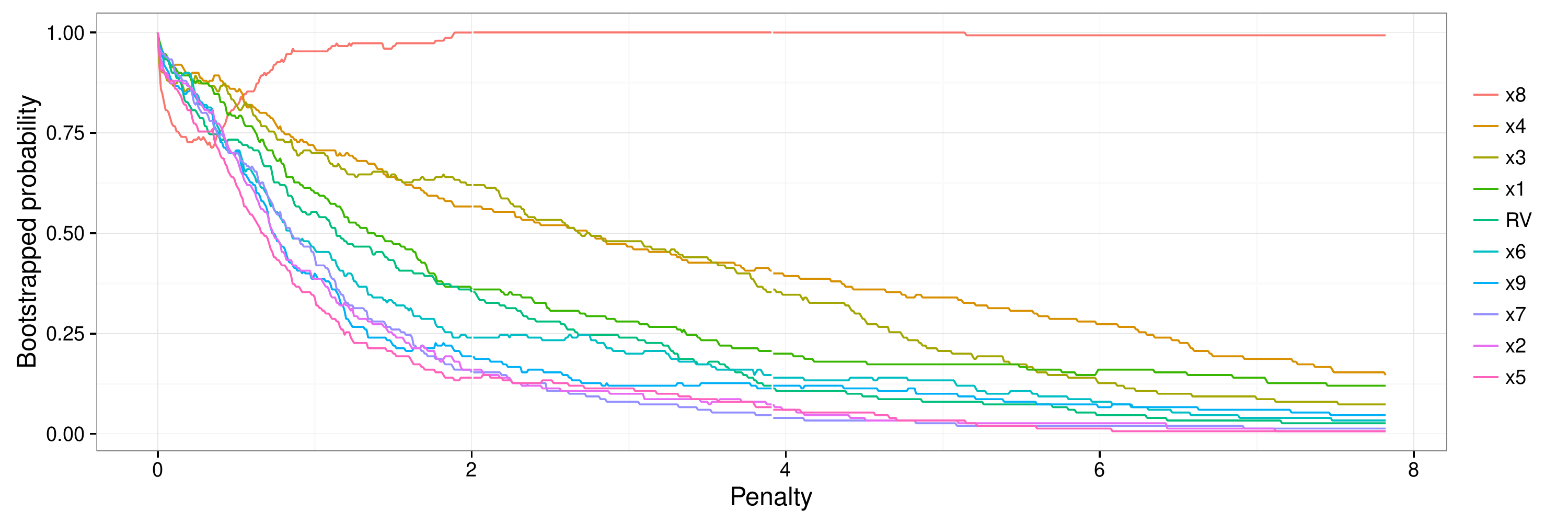}
\caption{Results of calls to \code{plot(vis.art, interactive = FALSE)} with additional arguments \code{which = "lvk"} in the top left, \code{which = "boot"} in the top right and \code{which = "vip"} down the bottom.}
\label{plot.vis}
\end{figure}

The \code{B = 150} argument provided to the \code{vis()} function tells us that we want to perform 150 bootstrap replications.  See \citet{Murray:2013} for more detail on the use of exponential weights in bootstrap model selection.    Specifying \code{redundant = TRUE} is unnecessary, as it is the default option; it ensures that an extra variable, randomly generated from a standard normal distribution and hence completely unrelated to the true data generating process, is added to the full model.  This extra redundant variable can be used as a baseline comparison in the variable inclusion plots. Finally, the \code{nbest} argument controls how many models with the smallest $\hat{Q}(\alpha)$ for each model size $k=1,\ldots,p$ are recorded.  It can take an integer argument or specifying \code{nbest = "all"} ensures that all possible models are displayed when the plot methods is called, as shown in the top left panel of Figure~\ref{plot.vis}. Typically researchers do not need to visualise the entire model space and in problems with larger numbers of candidate variables it is impractical to store and plot results for all models.  The default behaviour of the \code{vis()} function is to set \code{nbest = 5}, essentially highlighting the maximum enveloping lower convex curve of \citet{Murray:2013}.    

The simplest visualisation of the model space is to plot a measure of description loss against model complexity for all possible models, a special implementation is the Mallows $C_p$ plot \citep{Mallows:2000}. This is done using the argument \code{which = "lvk"} to the plot function applied to a \code{vis} object.  The string \code{"lvk"} is short for loss versus $k$, the dimension of the model.
\begin{CodeChunk}
\begin{CodeInput}
R> plot(vis.art, interactive = FALSE, highlight = "x8", which = "lvk")
\end{CodeInput}
\end{CodeChunk}
The result of this function can be found in the top left panel of Figure~\ref{plot.vis}.  The \code{highlight} argument is used to differentiate models that contain a particular variable from those that do not.  This is an implementation of the ``enriched scatter plot'' of \citet{Murray:2013}.  There is a clear separation between models that contain $x_8$ and those that do not, that is, all triangles are clustered towards the bottom with the circles above in a separate cluster.  There is no similar separation for the other explanatory variables (not shown).  These results strongly suggest that $x_8$ is the single most important variable.  For clarity the points have been jittered slightly along the horizontal axis, though the model sizes remain clearly differentiated. 

Rather than performing a single pass over the model space and plotting the description loss against model size, a more nuanced and discerning approach is to use a (exponential weighted) bootstrap to determine how often various models achieve the minimal loss for each model size.  The advantage of the bootstrap approach is that it gives a measure of model stability for each model size as promoted by  \citet{Meinshausen:2010}, \citet{Mueller:2010} and \citet{Murray:2013}.  

The weighted bootstrap has two key-benefits over the residual or nonparametric bootstrap: First, the weighted bootstrap always yields observable responses which is particularly relevant when these observable values are restricted to be integers (as in many generalized linear models), or, when $y$ values are naturally bounded, say to be observed on the interval 0 to 1; Second, the weighted bootstrap does not suffer from separation issues that regularly occur in logistic and other models.  The pairs bootstrap also yields observable responses and can be thought of as a special (boundary) case of the weighted bootstrap where some weights are allowed to be exactly zero, which can create a separation issue in logistic models.  Therefore, we have chosen to implement the weighted bootstrap because it is a simple, elegant method that appears to work well.  Specifically, we utilise the exponential weighted bootstrap where the observations are reweighted with weights drawn from an exponential distribution with mean 1  (see  \citet{Murray:2013} for more detail).

To visualise the results of the exponential weighted bootstrap, the \code{which = "boot"} argument needs to be passed to the plot call on a \code{vis} object.  The \code{highlight} argument can again be used to distinguish between models with and without a particular variable.  Each circle represents a model with a non-zero bootstrap probability, that is, each model that was selected as the best model of a particular dimension in at least one bootstrap replication.  Furthermore, the area of each circle is proportional to the corresponding model's bootstrapped selection probability.  

Figure~\ref{plot.vis} is an example of a model stability plot for the artificial data set.  The null model, the full model and the simple linear regression of $y$ on $x_8$ all have bootstrap probabilities equal to one. While there are alternatives to the null and full model their inclusion in the plot serves two main purposes.  Firstly, to gauge the potential range in description loss and secondly to provide a baseline against which to compare other circles to see if any approach a similar size, which would indicate that those are dominant models of a given model dimension.  In Figure~\ref{plot.vis}, for model dimensions of between three and ten, there are no clearly dominant models, that is, within each model size there are no models that are selected much more commonly than the alternatives. 

A print method is available for \code{vis} objects which prints the model formula, log-likelihood and proportion of times that a given model was selected as the ``best'' model within each model size.  The default minimum probability of a model being selected before it gets printed is 0.3, though this can be customised by passing a \code{min.prob} argument to the \code{print} function.
\begin{CodeChunk}
\begin{CodeInput}
R> print(vis.art, min.prob = 0.25)
\end{CodeInput}
\begin{CodeOutput}
                         name prob logLikelihood
                          y~1 1.00       -135.33
                         y~x8 1.00       -105.72
                      y~x4+x8 0.40       -103.63
                      y~x1+x8 0.27       -104.47
    y~x1+x2+x3+x4+x5+x6+x7+x9 0.26       -100.63
 y~x1+x2+x3+x4+x5+x6+x7+x9+RV 0.33       -100.51
\end{CodeOutput}
\end{CodeChunk}
The output above, reinforces what we know from the top right panel of Figure~\ref{plot.vis}.  The null model is always selected and in models of size two a regression of $y$ on $x_8$ is always selected.  In models of size three the two most commonly selected models are \code{y~x4+x8}, which was selected 40\% of the time and \code{y~x1+x8} selected in 27\% of bootstrap replications.  Interestingly, in models of size nine and ten, the most commonly selected models do not contain $x_8$, these are shown as blue circles in the plot.  We will see in the next section that this phenomenon is related to the failure of stepwise variable selection with this data set.  

\subsection{Variable inclusion plots}

Rather than visualising a loss measure against model size, it can be instructive to consider which variables are present in the overall ``best'' model over a set of bootstrap replications.  To facilitate comparison between models of different sizes we use the GIC, equation \eqref{GIC}, which includes a penalty term for the number of variables in each model.

Using the same exponential weighted bootstrap replications as in the model selection plots, we have a set of $B$ bootstrap replications and for each model size we know which model has the smallest description loss.   This information is used to determine which model minimises the GIC over a range of values of the penalty parameter, $\lambda$, in each bootstrap sample.  For each value of $\lambda$, we extract the variables present in the ``best'' models over the $B$ bootstrap replications and calculate the corresponding bootstrap probabilities that a given variable is present.  These calculations are visualised in a variable inclusion plot (VIP) as introduced by \citet{Mueller:2010} and \citet{Murray:2013}. The VIP shows empirical inclusion probabilities as a function of the penalty multiplier $\lambda$. The probabilities are calculated by observing how often each variable is retained in $B$ exponential weighted bootstrap replications.  Specifically, for each bootstrap sample $b=1,\ldots,B$ and each penalty multiplier $\lambda$, the chosen model, $\hat{\alpha}_{\lambda}^{b}\in \mathcal{A}$, is that which achieves the smallest $\textrm{GIC}(\alpha,\lambda;\mathbf{w}_b) = \hat{Q}^b(\alpha)+\lambda p_{\alpha}$, where $\mathbf{w}_b$ is the $n$-vector of independent and identically distributed exponential weights (we refer to Section 2.5 in \citet{Murray:2013} for more information on the weighted bootstrap). The inclusion probability for variable $x_{j}$ is estimated by $B^{-1}\sum_{i=1}^{B}\mathbb{I}\{j\in \hat{\alpha}_{\lambda}^{b}\}$, where $\mathbb{I}\{j\in \hat{\alpha}_{\lambda}^{b}\}$ is one if $x_{j}$ is in the final model and zero otherwise.  Following \citet{Murray:2013}, the default range of $\lambda$ values is $\lambda\in[0,2\log(n)]$ as this includes most standard values used for the penalty parameter.

The example shown in the bottom panel of Figure~\ref{plot.vis} is obtained using the \code{which = "vip"} argument to the plot function.  As expected, when the penalty parameter is equal to zero, all variables are included in the model;  the full model achieves the lowest description loss, and hence minimises the GIC when there is no penalisation.  As the penalty parameter increases, the inclusion probabilities for individual variables typically decrease as more parsimonious models are preferred.  In the present example, the inclusion probabilities for the $x_8$ variable exhibit a sharp decrease at low levels of the penalty parameter, but then increase steadily as a more parsimonious model is sought.  This pattern helps to explain why stepwise model selection chose the larger model with all the variables except $x_8$ -- there exists a local minimum.  Hence, for large models the inclusion of $x_8$ adds no additional value over having all the other explanatory variables in the model.

It is often instructive to visualise how the inclusion probabilities change over the range of penalty parameters.  The ordering of the variables in the legend corresponds to their average inclusion probability over the whole range of penalty values.  We have also added an independent standard Gaussian random variable to the model matrix as a redundant variable (\code{RV}).  This provides a baseline to help determine which inclusion probabilities are ``significant'' in the sense that they exhibit a different behaviour to the \code{RV} curve.  Variables with inclusion probabilities near or below the \code{RV} curve can be considered to have been included by chance.  

To summarise, VIPs continue the model stability theme. Rather than simply using a single penalty parameter associated with a particular information criterion, for example the AIC with $\lambda=2$, our implementation of VIPs adds considerable value by allowing us to learn from a range of penalty parameters.  Furthermore, we are able to see which variables are most often included over a number of bootstrap samples.  An alternative approach to assessing model stability, the simplified adaptive fence, is introduced in the next section.

\section{The simplified adaptive fence}\label{sec:af}

The fence, first introduced by \citet{Jiang:2008}, is built around the inequality 
$$\hat{Q}(\alpha) - \hat{Q}(\alpha_{f}) \leq c,$$ 
where $\hat Q$ is an empirical measure of description loss, $\alpha$ is a candidate model and $\alpha_{f}$ is the baseline, ``full'' model.  The procedure attempts to isolate a set of ``correct models'' that satisfy the inequality.  A model $\alpha^*$, is described as ``within the fence'' if $\hat{Q}(\alpha^*) - \hat{Q}(\alpha_{f}) \leq c$. From the set of models within the fence, the one with minimum dimension is considered optimal. If there are multiple models within the fence at the minimum dimension, then the model with the smallest $\hat{Q}(\alpha)$ is selected.  For a recent review of the fence and related methods, see \citet{Jiang:2014}.

The implementation we provide in the \pkg{mplot} package is inspired by the simplified adaptive fence proposed by \citet{Jiang:2009}, which represents a significant advance over the original fence method proposed by \citet{Jiang:2008}.  The key difference is that the parameter $c$ is not fixed at a certain value, but is instead adaptively chosen.  Simulation results have shown that the adaptive method improves the finite sample performance of the fence, see  \citet{Jiang:2008,Jiang:2009}.

The adaptive fence procedure entails bootstrapping over a range of values of the parameter $c$.  For each value of $c$ a parametric bootstrap is performed under $\alpha_f$.  For each bootstrap sample we identify the smallest model inside the fence, $\hat{\alpha}(c)$. \citet{Jiang:2009} suggest that if there is more than one model, choose the one with the smallest $\hat{Q}(\alpha)$.  Define the  empirical probability of selecting model $\alpha$ for a given value of $c$ as $p^*(c,\alpha)=P^*\{\hat{\alpha}(c)=\alpha\}$.  Hence, if $B$ bootstrap replications are performed, $p^*(c,\alpha)$ is the proportion of times that model $\alpha$ is selected.  Finally, define an overall selection probability, $p^*(c)=\max_{\alpha\in\mathcal{A}}p^*(c,\alpha)$ and plot $p^*(c)$ against $c$ to find the first peak. The value of $c$ at the first peak, $c^*$, is then used with the standard fence procedure on the original data.

\begin{figure}
\includegraphics[width=0.5\textwidth]{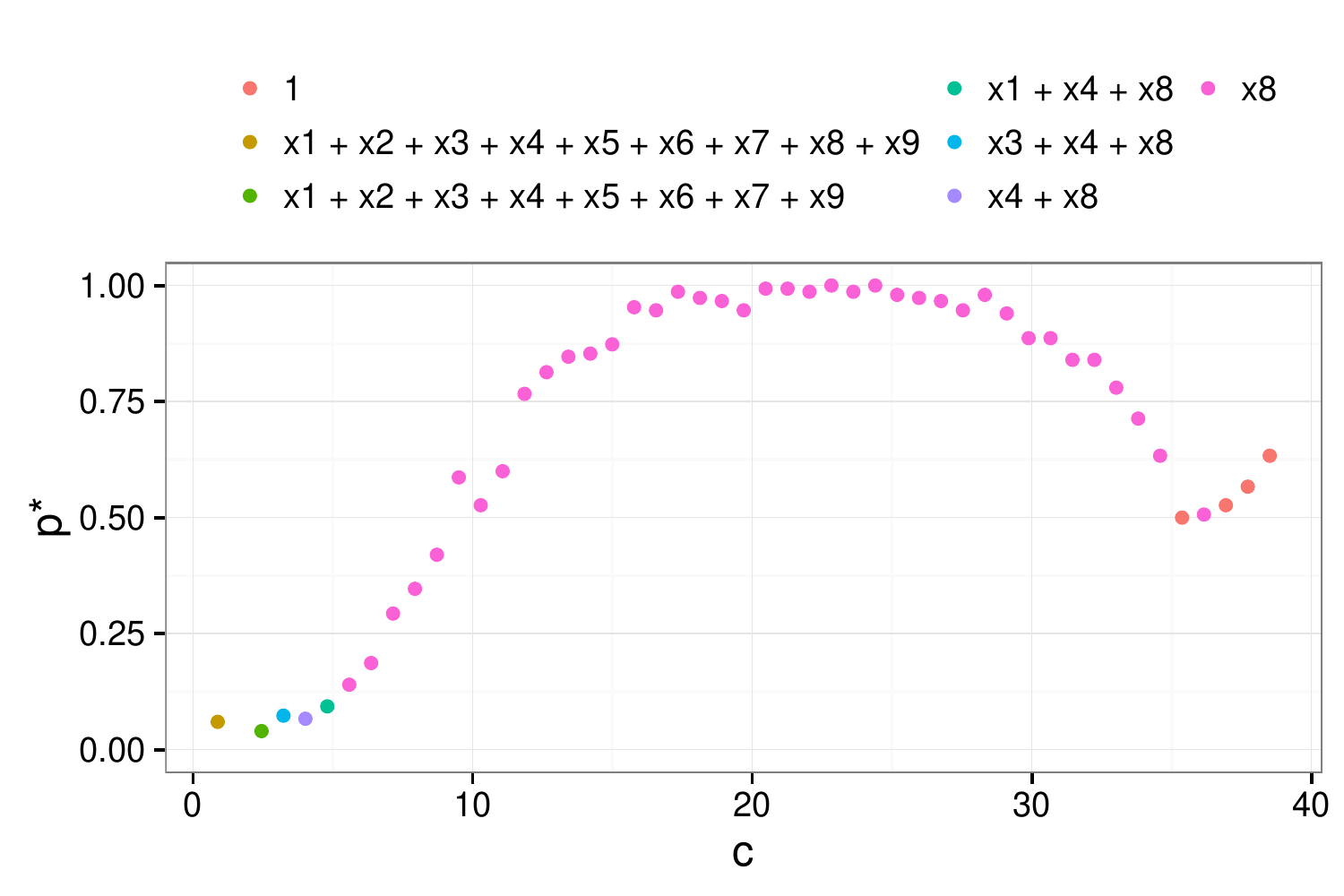}\includegraphics[width=0.5\textwidth]{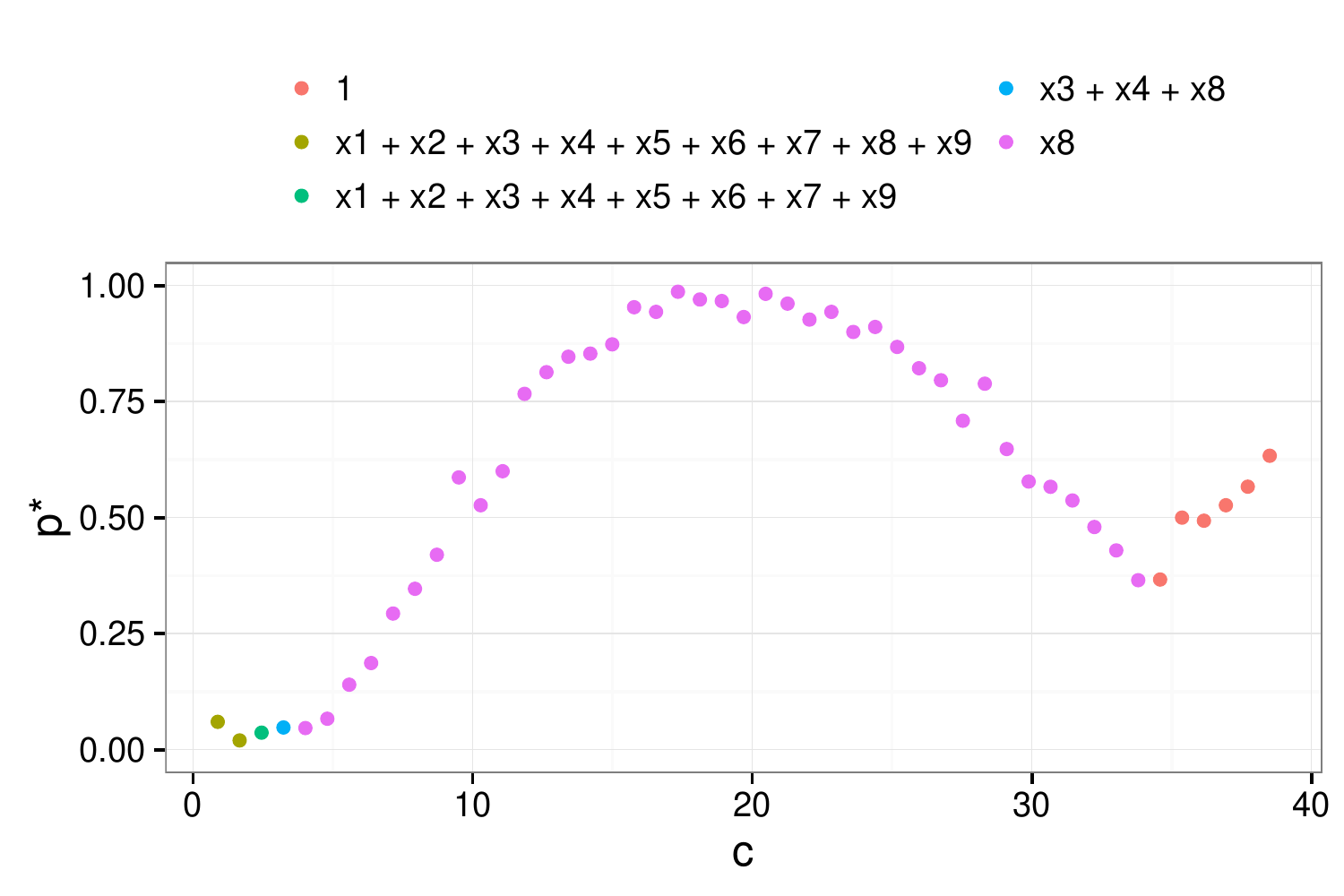}
\caption{Result of a call to \code{plot(af.art, interactive = FALSE)} with additional arguments \code{best.only = TRUE} on the left and \code{best.only = FALSE} on the right.  The more rapid decay after the $x_8$ model is typical of using \code{best.only = FALSE} where the troughs between candidate/dominant models are more pronounced.}
\label{plot.af}
\end{figure}

Our implementation is provided through the \code{af()} function and associated plot methods. An example with the artificial data set is given in Figure~\ref{plot.af} which is generated using the following code.
\begin{CodeChunk}
\begin{CodeInput}
R> af.art = af(lm.art, B = 150, n.c = 50)
R> plot(af.art, interactive = FALSE, best.only = TRUE)
\end{CodeInput}
\end{CodeChunk} 
The arguments indicate that we perform $B = 150$ bootstrap resamples, over a grid of $50$ values of the parameter $c$. In this example, there is only one peak, and the choice of $c^*=21.1$ is clear.

One might expect that there should be a peak corresponding to the full model at $c=0$, but this is avoided by the inclusion of at least one redundant variable.  Any model that includes the redundant variable is known to not be a ``true'' model and hence is not included in the calculation of $p^*(c)$.  This issue was first identified and addressed by \citet{Jiang:2009}.

There are a number of key differences between our implementation and the method proposed by \citet{Jiang:2009}.  Perhaps the most fundamental difference is in the philosophy underlying our implementation.  Our approach is more closely aligned with the concept of model stability than with trying to pick a single ``best'' model.  This can be seen through the plot methods we provide.  Instead of simply using the plots to identify the first peak, we add a legend that highlights which models were the most frequently selected for each parameter value, that is, for each $c$ value we identify which model gave rise to the $p^*(c)$ value.   In this way, researchers can ascertain if there are regions of stability for various models.  In the example given in Figure~\ref{plot.af}, there is no need to even define a $c^*$ value, it is obvious from the plot that there is only one viable candidate model, a regression of $y$ on $x_8$.

Our approach considers not just the best model of a given model size, but also allows users to view a plot that takes into account the possibility that more than one model of a given model size is within the fence.  The \code{best.only = FALSE} option when plotting the results of the adaptive fence is a modification of the adaptive fence procedure which considers all models of a particular size that are within the fence when calculating the $p^*(c)$ values. In particular, for each value of $c$ and for each bootstrap replication, if a candidate model is found inside the fence, then we look to see if there are any other models of the same size that are also within the fence. If no other models of the same size are inside the fence, then that model is allocated a weight of 1. If there are two models inside the fence, then the best model is allocated a weight of 1/2. If three models are inside the fence, the best model gets a weight of 1/3, and so on. After $B$ bootstrap replications, we aggregate the weights by summing over the various models. The $p^*(c)$ value is the maximum aggregated weight divided by the number of bootstrap replications. This correction penalises the probability associated with the best model if there were other models of the same size inside the fence. The rationale is that if a model has no redundant variables then it will be the only model of that size inside the fence over a range of values of $c$. The result is more pronounced peaks which can help to determine the location of the correct peak and identify the optimal $c^*$ value or more clearly differentiate regions of model stability.  This can be seen in the right hand panel of Figure~\ref{plot.af}.

Another key difference is that our implementation is designed for linear and generalised linear models, rather than mixed models.  As far as we are aware, this is the first time fence methods have been applied to such models.  There is potential to add mixed model capabilities to future versions of the \pkg{mplot} package, but computational speed is a major hurdle that needs to be overcome.  The current implementation is made computationally feasible through the use of the \pkg{leaps} and \pkg{bestglm} packages and the use of parallel processing, as discussed in Section \ref{sec:timing} \citep{Lumley:2009,McLeod:2014}.

We have also provided an optional initial stepwise screening method that can help limit the range of $c$ values over which to perform the adaptive fence procedure.  The initial stepwise procedure performs forward and backward stepwise model selection using both the AIC and BIC.  From the four candidate models, we extract the size of smallest and largest models, $k_L$ and $k_U$ respectively.  To obtain a sensible range of $c$ values we consider the set of models with dimension between $k_L-2$ and $k_U+2$.  Due to the inherent limitations of stepwise procedures, outlined in Section \ref{sec:ie}, it can be useful to check \code{initial.stepwise = FALSE} with a small number of bootstrap replications over a sparse grid of $c$ values to ensure that the \code{initial.stepwise = TRUE} has produced a reasonable region.

\section{Interactive graphics}\label{sec:ig}

To facilitate that researchers can more easily gain value from the static plots given in Figures~\ref{plot.vis} and \ref{plot.af} and to help them interact with the model selection problem more closely, we have provided a set of interactive graphics based on the \pkg{googleVis} package and wrapped them in a \pkg{shiny} user interface.  It is still quite novel for a package to provide a shiny interface for its methods, but there is precedent, see, for example \citet{McMurdie:2013} or \citet{Gabry:2015}.

Among the most important contributions of these interactive methods is: the provision of tooltips to identify the models and/or variables; pagination of the legend for the variable inclusion plots; and a way to quickly select which variable to highlight in the model stability plots. These interactive plots can be generated when the \code{plot()} function is run on an \code{af} or \code{vis} object by specifying \code{interactive=TRUE}.

\begin{figure}
\centering
\includegraphics[width=0.95\textwidth, clip, trim = 57 84 70 130]{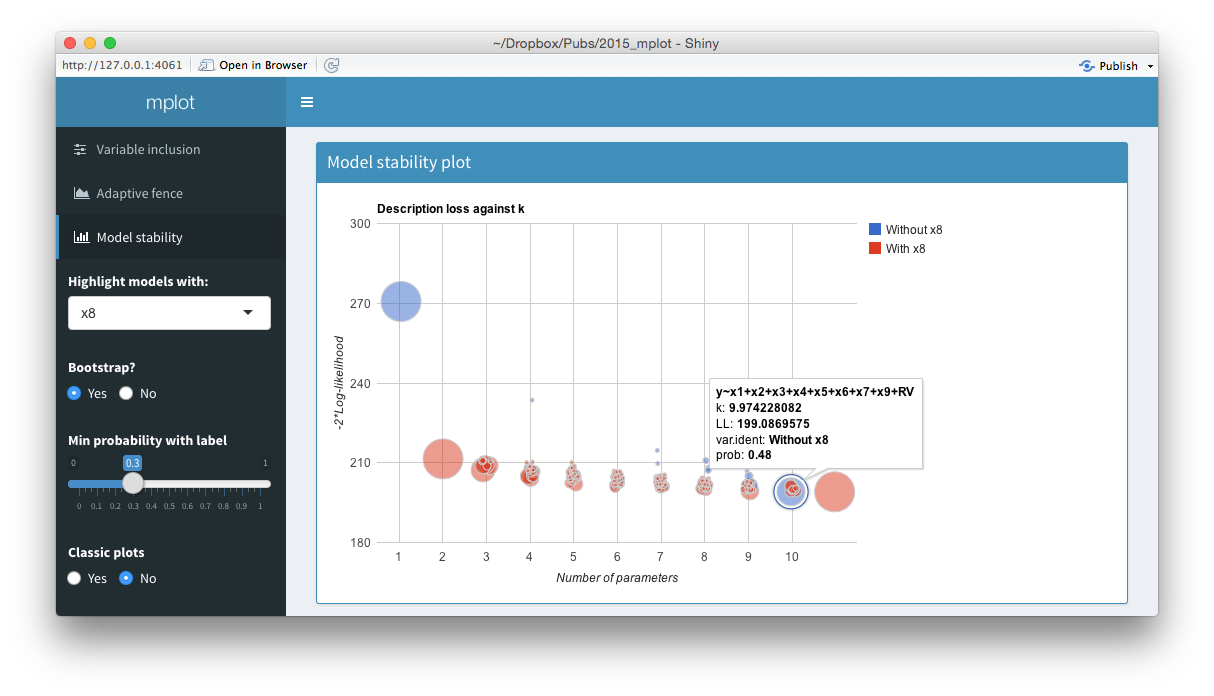}

\vspace{2mm}

\includegraphics[width=0.95\textwidth, clip, trim = 57 84 70 130]{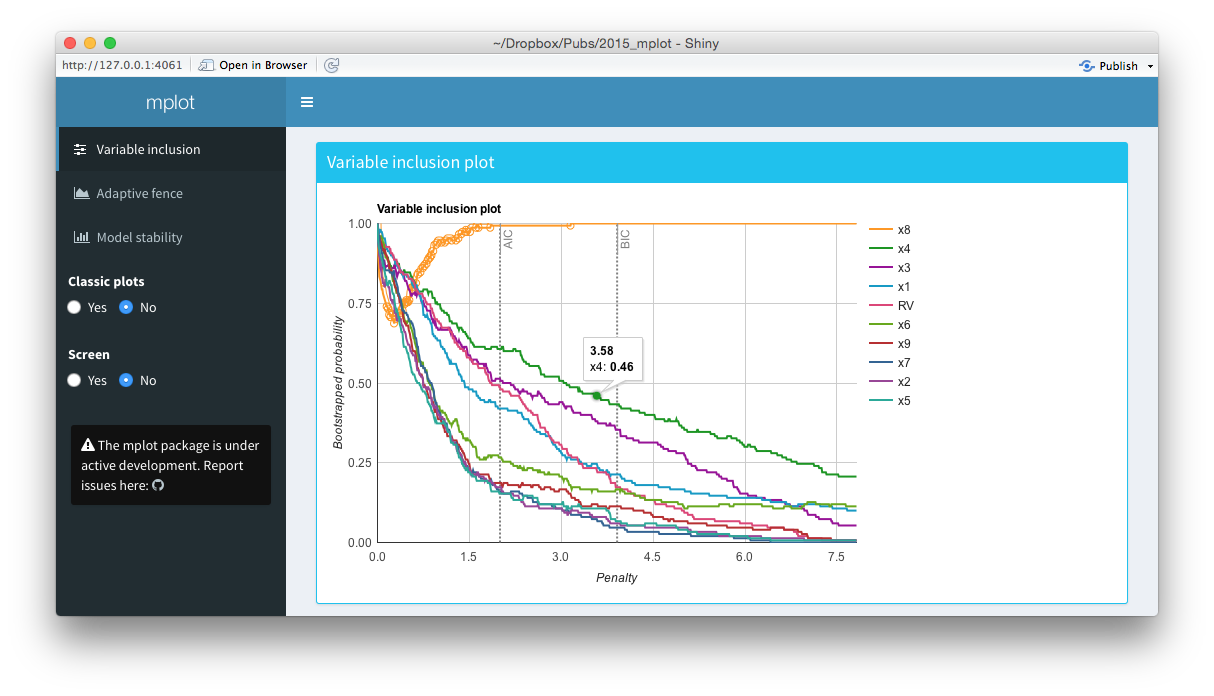}

\vspace{2mm}

\includegraphics[width=0.95\textwidth, clip, trim = 57 84 70 130]{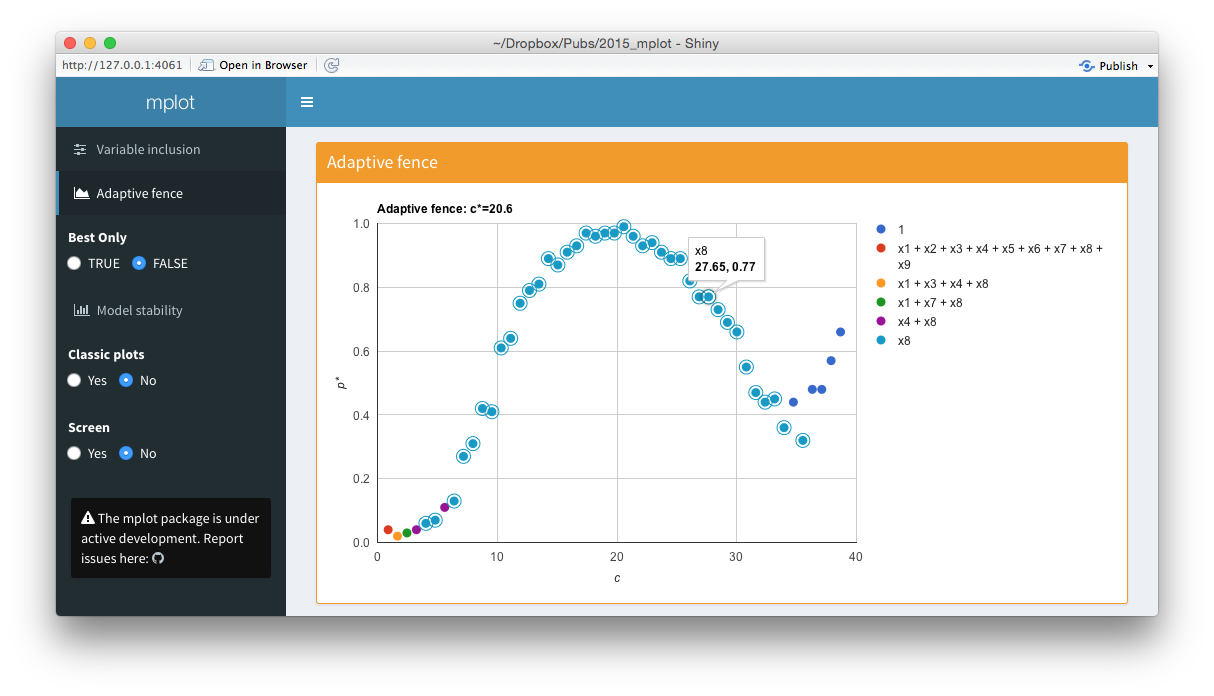}
\caption{Screenshots from the web interface generated using \code{mplot()}.}
\label{fig:shiny}
\end{figure}

The \pkg{mplot} package takes interactivity a step further, embedding these plots within a shiny web interface.  This is done through a call to the \code{mplot()} function, which requires the full fitted model as the first argument and then a \code{vis} object and/or \code{af} object (in any order).
\begin{CodeChunk}
\begin{CodeInput}
R> mplot(lm.art, vis.art, af.art)
\end{CodeInput}
\end{CodeChunk}
Note that the \code{vis()} and \code{af()} functions need to be run and the results stored prior to calling the \code{mplot()} function.  The result of a call to this function is a webpage built using the \pkg{shiny} package with \pkg{shinydashboard} stylings \citep{Chang:2015a,Chang:2015b}.  Figure \ref{fig:shiny} shows a series of screen shots for the artificial example, equivalent to Figures \ref{plot.vis} and \ref{plot.af}, resulting from the above call to \code{mplot()}.  

The top panel of Figure \ref{fig:shiny} shows a model stability plot where the full model that does not contain $x_8$ has been selected and a tooltip has been displayed.  It gives details about the model specification,  the log-likelihood and the bootstrap selection probability within models of size 10.  The tooltip makes it  easier for users to identify which variables are included in dominant models than the static plot equivalent.  On the left hand side of the shiny interface, a drop down menu allows users to select the variable to be highlighted.  This is passed through the \code{highlight} argument discussed in Section \ref{sec:msp}.  Models with the highlighted variable are displayed as red circles whereas models without the highlighted variable are displayed as blue circles.  The ability for researchers to quickly and easily see which models in the stability plot contain certain variables enhances their understanding of the relative importance of different components in the model.  Selecting ``No'' at the ``Bootstrap?'' radio buttons yields the plot of description loss against dimension shown in the top left panel of Figure \ref{plot.vis}.

The middle panel of Figure \ref{fig:shiny} is a screen shot of an interactive variable inclusion plot. When the mouse hovers over a line, the tooltip gives information about the bootstrap inclusion probability and which variable the line represents.  Note that in comparison to the bottom panel of Figure \ref{plot.vis}, the legend is now positioned outside of the main plot area.  When the user clicks a variable in the legend, the corresponding line in the plot is highlighted.  This can be seen in Figure \ref{fig:shiny}, where the $x_8$ variable in the legend has been clicked and the corresponding $x_8$ line in the variable inclusion plot has been highlighted.  The highlighting is particularly useful with the redundant variable, so it can easily be identified.  If the number of predictor variables is such that they no longer fit neatly down the right hand side of the plot, they simply paginate, that is an arrow appears allowing users to toggle through to the next page of variables.  This makes the interface cleaner and easier to interpret than the static plots.  Note also the vertical lines corresponding to traditional AIC and BIC penalty values.

The bottom panel of Figure \ref{fig:shiny} is an interactive adaptive fence plot. The tooltip for a particular point gives information about the explanatory variable(s) in the model, the $\alpha^*=\arg\max_{\alpha\in\mathcal{A}}p^*(c,\alpha)$ value and the $(c,p^*(c))$ pair that has been plotted.  Hovering or clicking on a model in the legend highlights all the points in the plot corresponding to that model.  In this example, the $x_8$ legend has been clicked on and an additional circle has been added around all points representing the regression with $x_8$ as the sole explanatory variable.  The shiny interface on the left allows users to toggle between \code{best.only = TRUE} and \code{best.only = FALSE}.

The interactive graphics and shiny interface are most useful in the exploratory stage of model selection.  Once the researcher has found the most informative plot through interactive analysis, the more traditional static plots may be used in a formal write up of the problem.

\section{Timing}\label{sec:timing}

Any bootstrap model selection procedure is time consuming.  However, for linear models, we have leveraged the efficiency of the branch-and-bound algorithm provided by \pkg{leaps} \citep{Miller:2002,Lumley:2009}.  The \pkg{bestglm} package is used for GLMs; but in the absence of a comparably efficient algorithm the computational burden is much greater \citep{McLeod:2014}.

Furthermore, we have taken advantage of the embarrassingly parallel nature of bootstrapping, utilising the \pkg{doParallel} and \pkg{foreach} packages to provide cross platform multicore support, available through the \code{cores} argument \citep{doParallel:2014,foreach:2014}.  By default it will detect the number of cores available on your computer and leave one free.

\begin{figure}[t]
\centering
\includegraphics[width=0.5\textwidth]{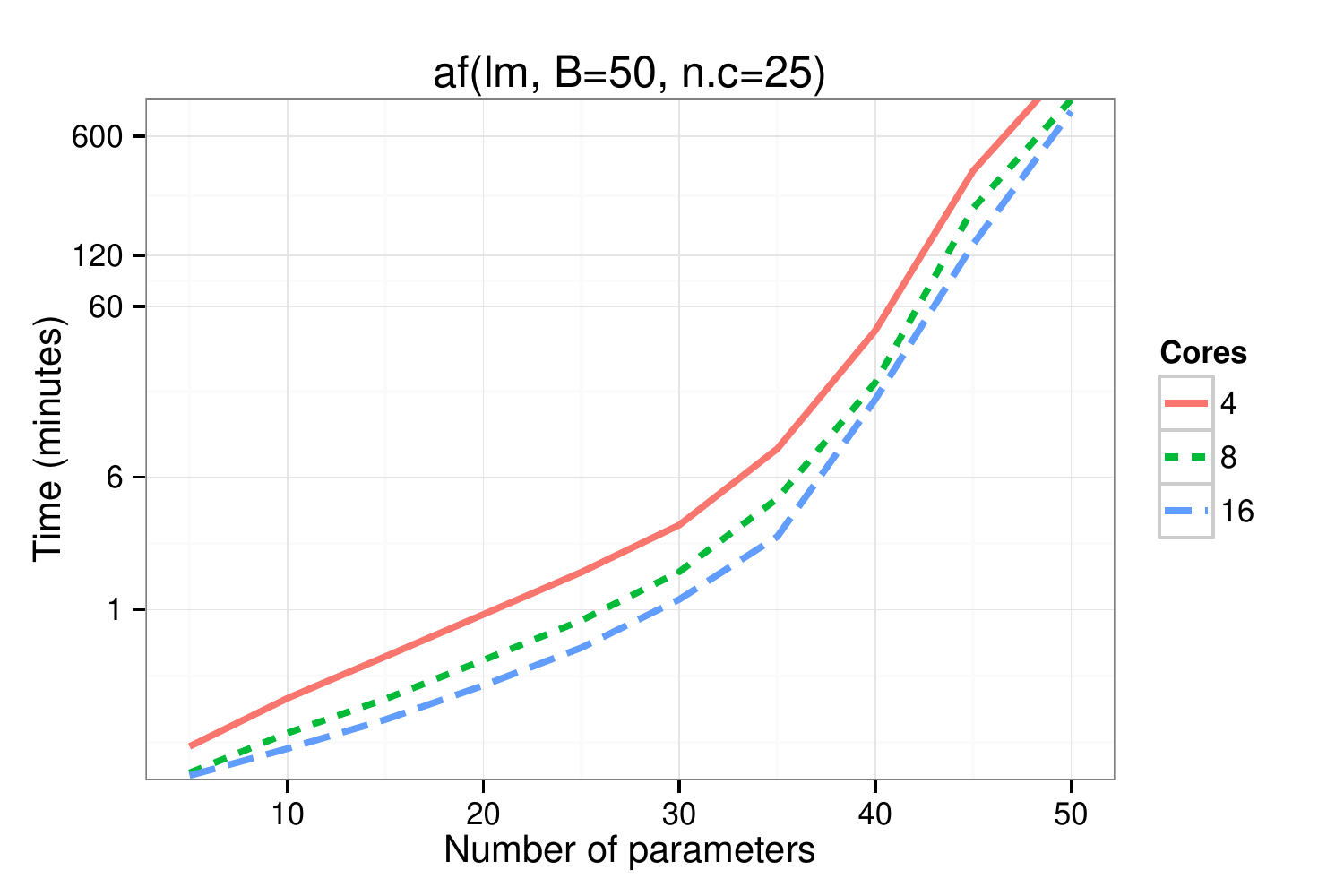}\includegraphics[width=0.5\textwidth]{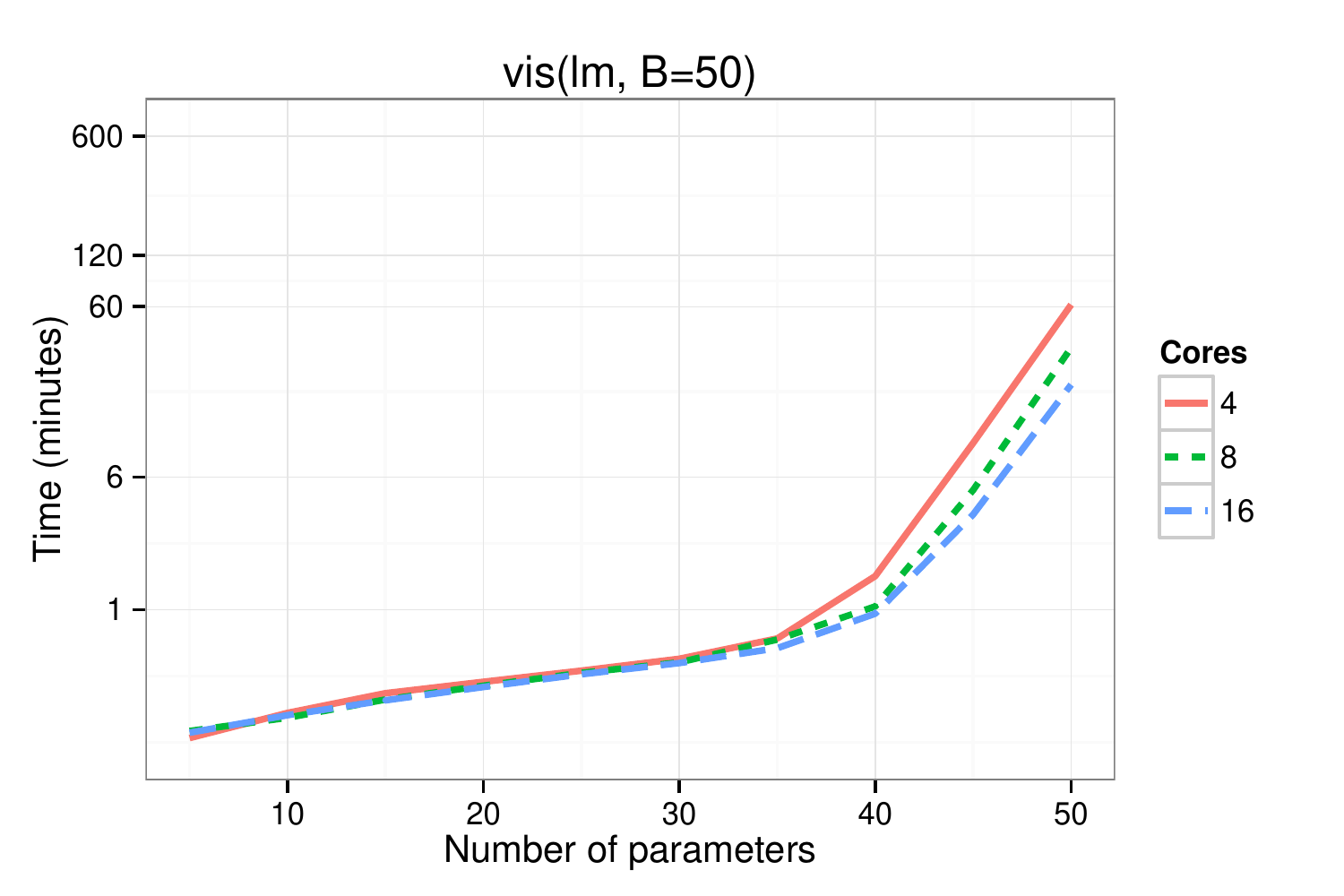}
\includegraphics[width=0.5\textwidth]{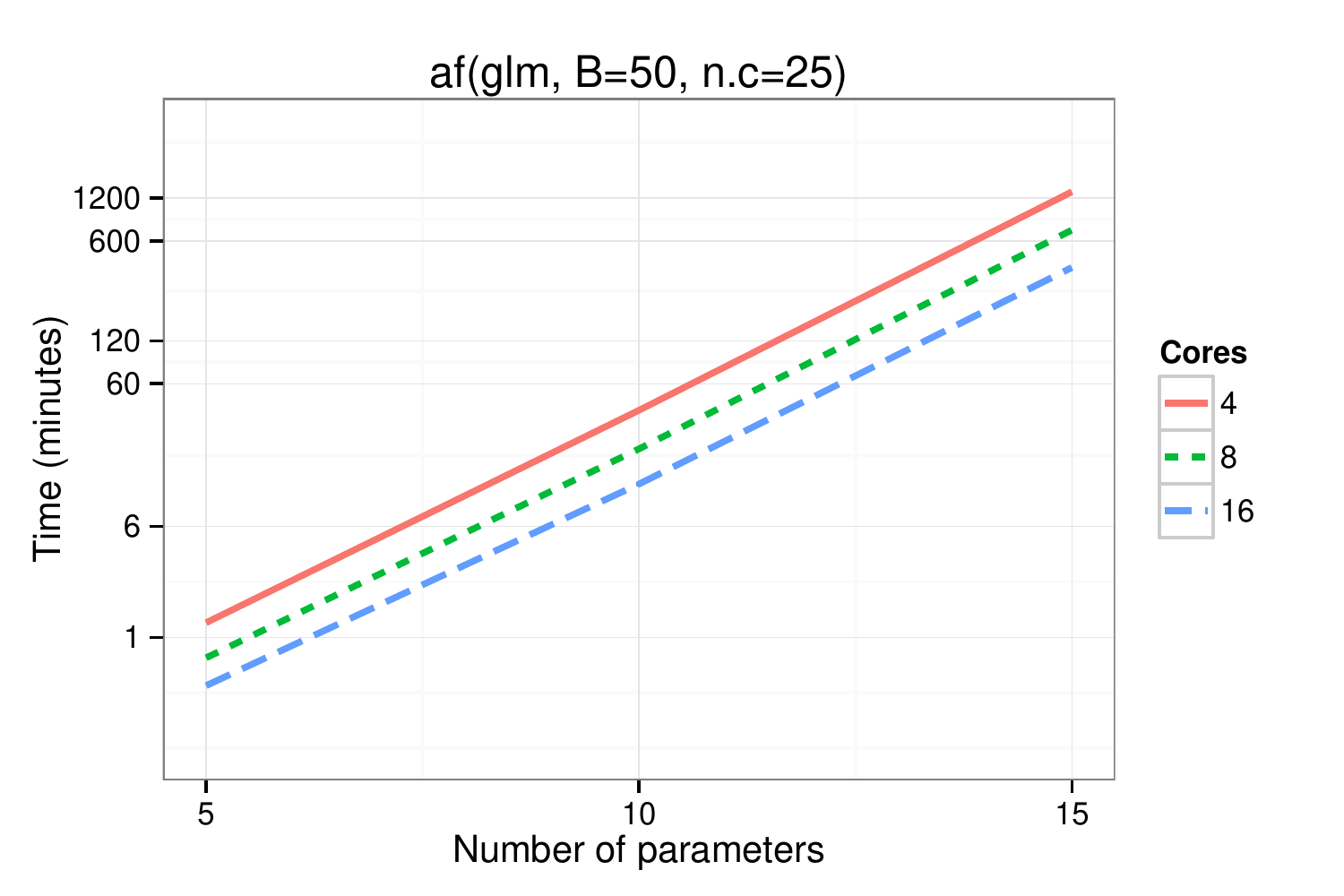}\includegraphics[width=0.5\textwidth]{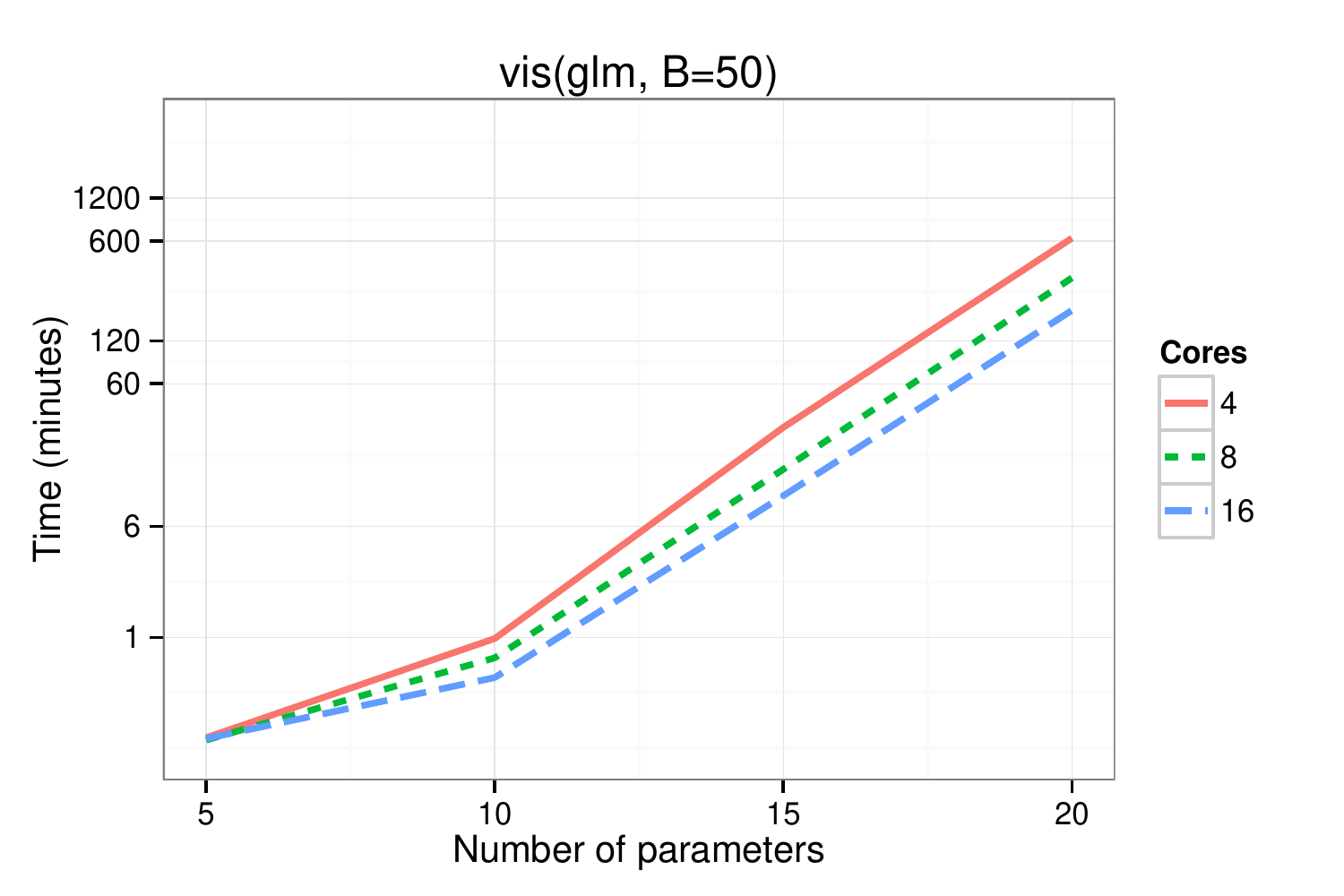}
\caption{Average time required to run the \code{af()} and \code{vis()} functions when $n=100$.  A binomial regression was used for the GLM example.}
\label{fig:time}
\end{figure}

Figure \ref{fig:time} shows the timing results of simulations run for standard use scenarios with 4, 8 or 16 cores used in parallel.  Each observation plotted is the average of four runs of a given model size. The simulated models had a sample size of $n=100$ with $5,10,\ldots,50$ candidate variables, of which 30\% were active in the true model.

The results show both the \code{vis()} and \code{af()} functions are quite feasible on standard desktop hardware with 4 cores even for moderate dimensions of up to 40 candidate variables.  The adaptive fence takes longer than the \code{vis()} function, though this is to be expected as the effective number of bootstrap replications is \code{B}$\times$\code{n.c}, where \code{n.c} is the number divisions in the grid of the parameter $c$.

The results for GLMs are far less impressive, even when the maximum dimension of a candidate solution is set to \code{nvmax = 10}.  In its current implementation, the adaptive fence is only really feasible for models of around 10 predictors and the \code{vis()} function for 15.  Future improvements could see approximations of the type outlined by \citet{Hosmer:1989} to bring the power of the linear model branch-and-bound algorithm to GLMs.  An example of how this works in practice is given in Section \ref{sec:bw}.

An alternative approach for high dimensional models would be to consider subset selection with convex relaxations as in \citet{Shen:2012} or combine bootstrap model selection with regularisation.  In particular, we have implemented variable inclusion plots and model stability plots for \pkg{glmnet} \citep{Friedman:2010}.  In general, this is very fast for models of moderate dimension, but it does not consider the full model space.  Restrictions within the \pkg{glmnet} package, mean it is only applicable to linear models, binomial logistic regression, and Poisson regression with the log link function.  The \pkg{glmnet} package also allows for \code{"multinomial"}, \code{"cox"}, and \code{"mgaussian"} families, though we have not yet incorporated these into the \pkg{mplot} package.

\section{Real examples}\label{sec:examples}

\subsection{Diabetes example}\label{sec:diabetes}

Table \ref{tab:diabetes} shows a subset of the diabetes data used in \citet{Efron:2004}.  There are 10 explanatory variables, including age (\code{age}), sex (\code{sex}), body mass index (\code{bmi}) and mean arterial blood pressure (\code{map}) of 442 patients as well as six blood serum measurements (\code{tc}, \code{ldl}, \code{hdl}, \code{tch}, \code{ltg} and \code{glu}).  The response is a measure of disease progression one year after the baseline measurements. 

\begin{table}[t]
\centering
\begin{tabular*}{\textwidth}{@{\centering\extracolsep{\fill}}rcccR[.][]{3}{0}cR[.][.]{3}{1}ccccR[,][]{3}{0}}
\toprule
 &  &  &  & \multicolumn{1}{c}{} & \multicolumn{6}{c}{Serum measurements} & \multicolumn{1}{c}{Response} \\ \cmidrule(lr){6-11}
Patient &age&sex&bmi&\multicolumn{1}{c}{map}&tc&\multicolumn{1}{c}{ldl} & hdl & tch & ltg & glu & \multicolumn{1}{c}{$y$} \\ \midrule
1 		& 59 	& 2 	& 32.1 	& 101 	& 157 	& 93.2 & 38 & 4 & 4.9 & 87 & 151 \\
2 		& 48 	& 1 	& 21.6 	& 87 	& 183 	& 103.2 & 70 & 3 & 3.9 & 69 & 75 \\
3 		& 72 	& 2 	& 30.5 	& 93 	& 156 	& 93.6 & 41 & 4 & 4.7 & 85 & 141 \\
\vdots \ \ 	&\vdots &\vdots &\vdots &\multicolumn{1}{c}{\vdots} &\vdots &\multicolumn{1}{c}{\vdots} &\vdots &\vdots &\vdots &\vdots &\multicolumn{1}{c}{\vdots} \\
441 & 36 & 1 & 30.0 & 95 & 201 & 125.2 & 42 & 5 & 5.1 & 85 & 220 \\
442 & 36 & 1 & 19.6 & 71 & 250 & 133.2 & 97 & 3 & 4.6 & 92 & 57 \\ \bottomrule
\end{tabular*}
\caption{Measurements on 442 diabetes patients over 10 potential predictor variables and the response variable, a measure of disease progression after one year.}
\label{tab:diabetes}
\end{table} 

\begin{figure}[p]
\centering
\includegraphics[width=0.45\textwidth]{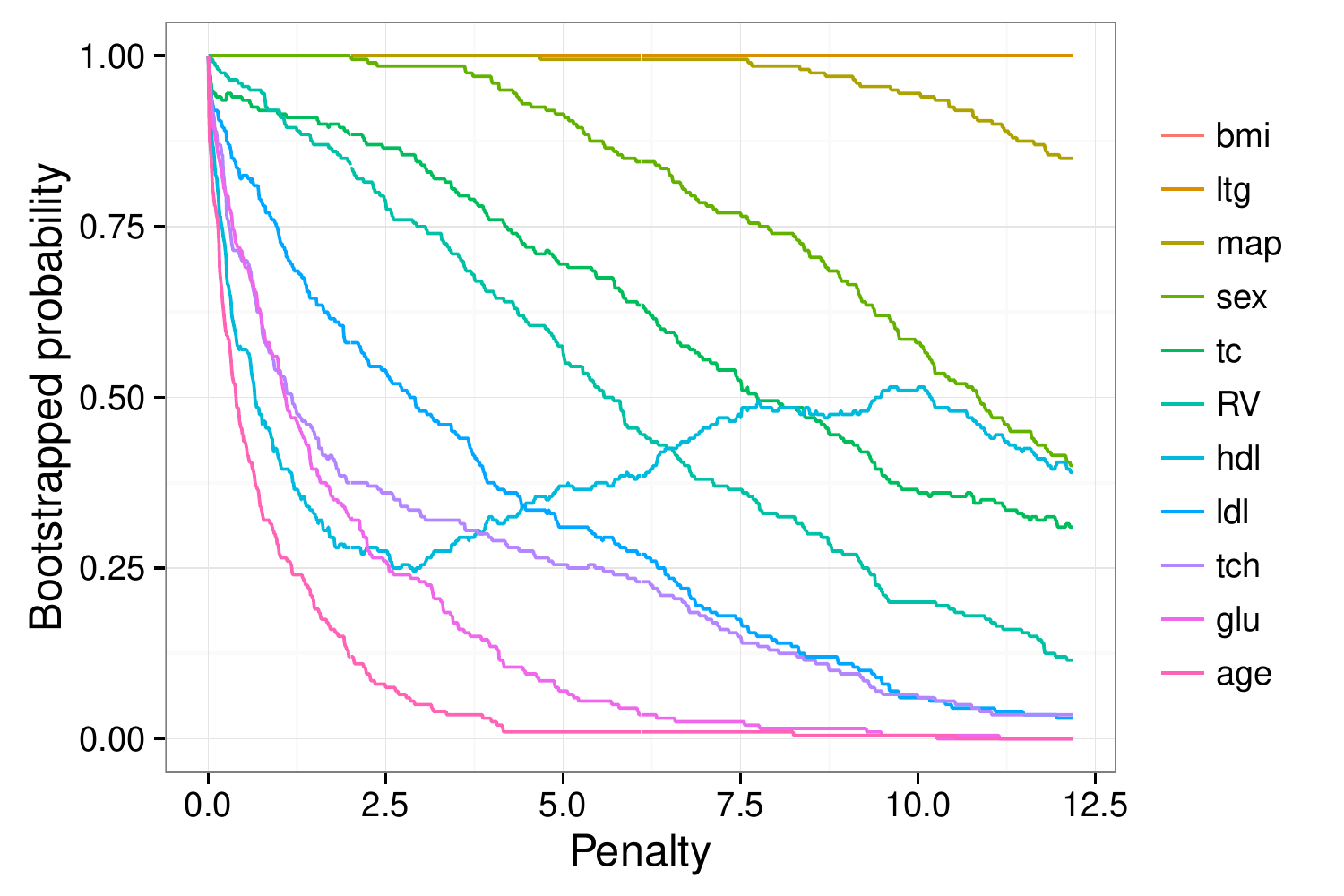}\includegraphics[width=0.45\textwidth]{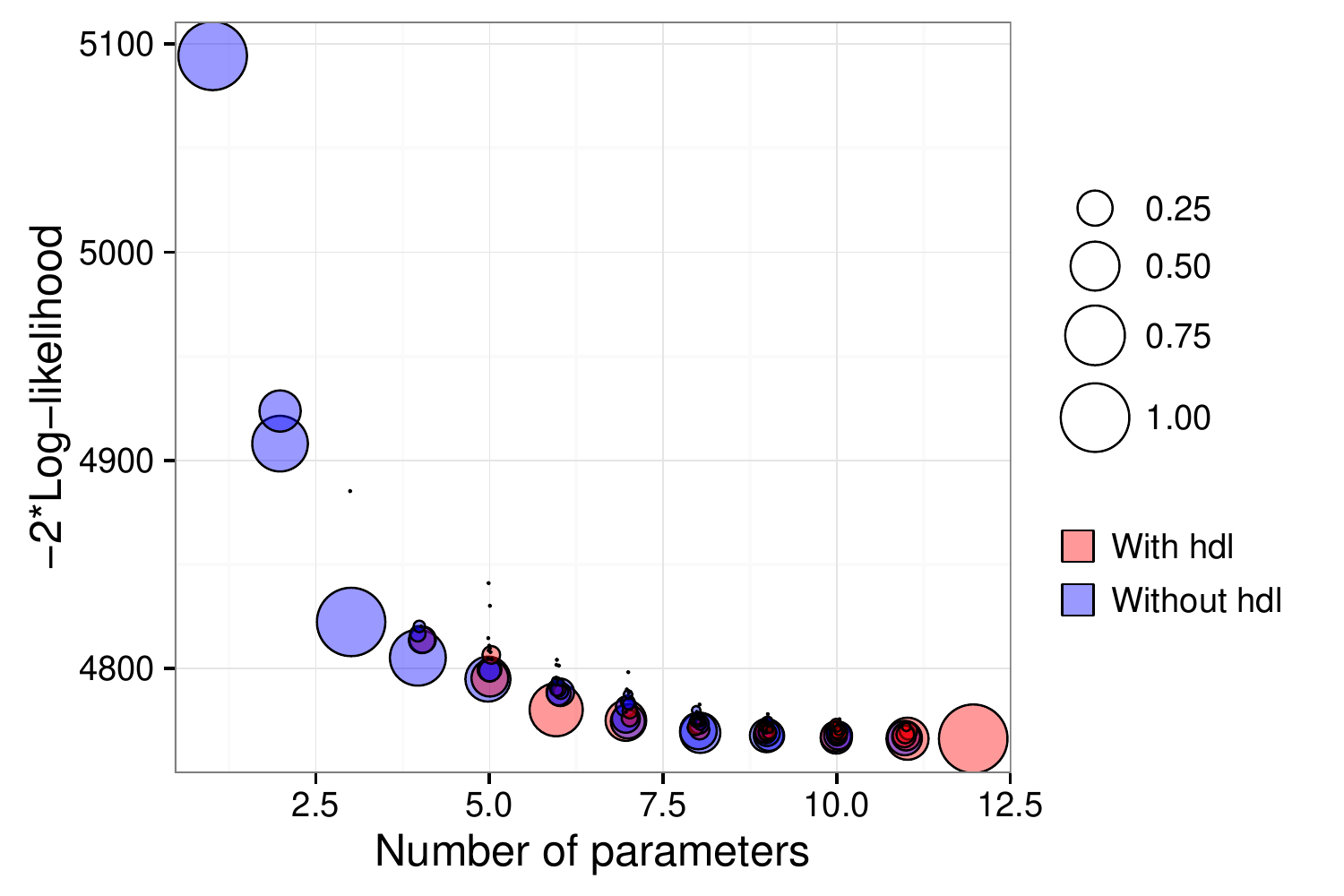}
\includegraphics[width=0.45\textwidth]{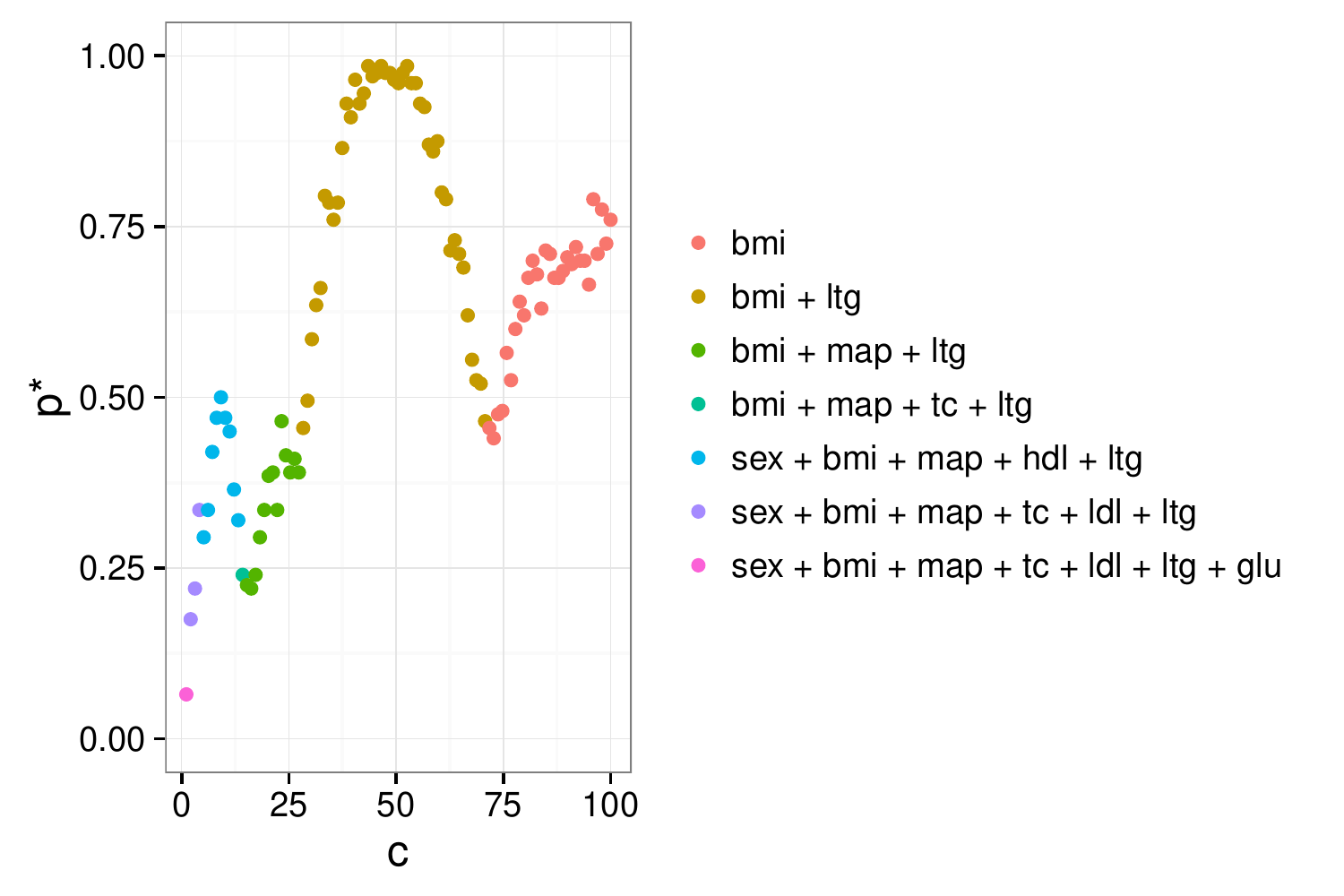}\includegraphics[width=0.45\textwidth]{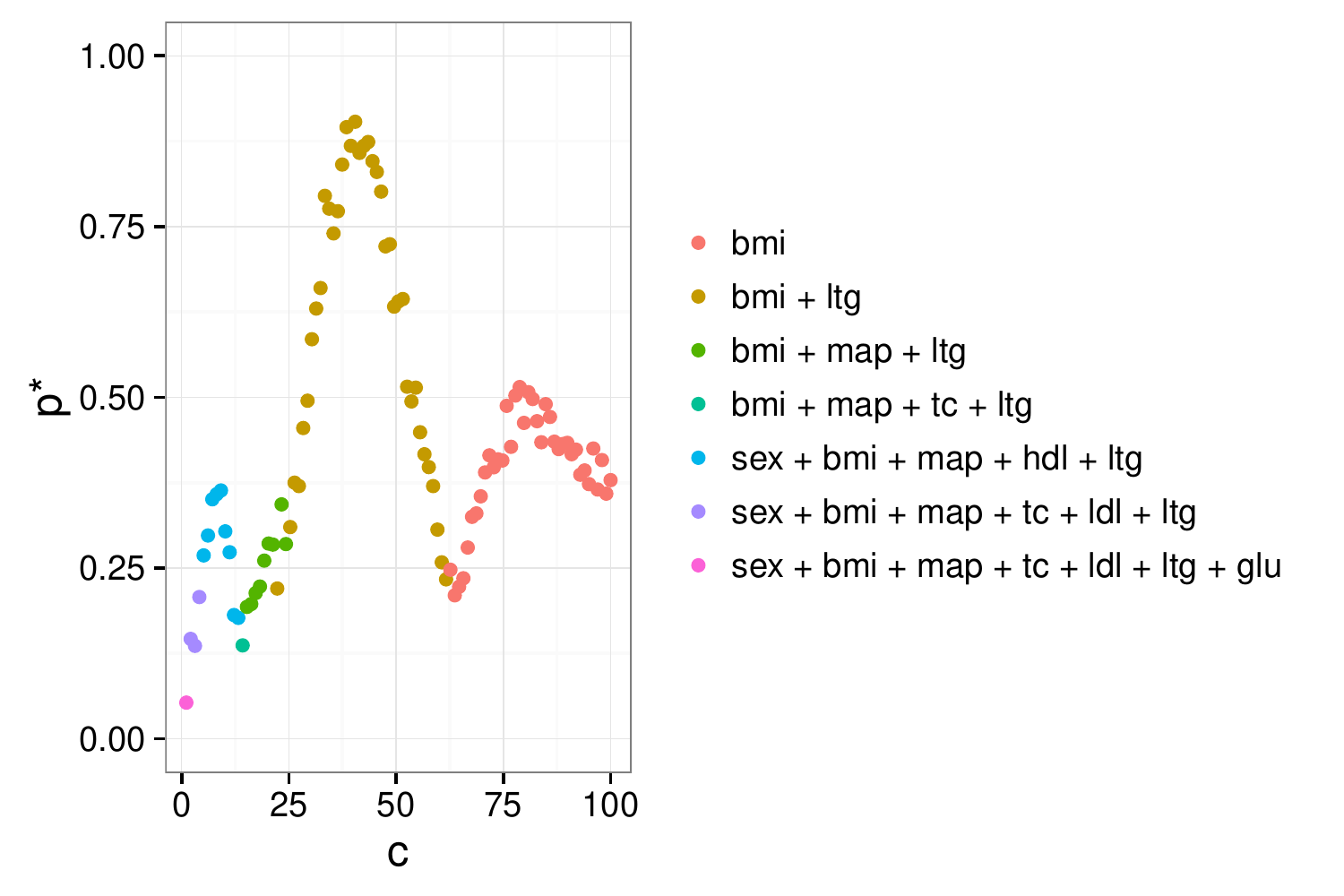}
\caption{Diabetes main effects example.}
\label{fig:diabetesmain}
\end{figure}

Figure \ref{fig:diabetesmain} shows the results of the main methods for the diabetes data obtained using the following code.

\begin{CodeChunk}
\begin{CodeInput}
R> lm.d = lm(y ~ ., data = diabetes)
R> vis.d = vis(lm.d, B = 200)
R> af.d = af(lm.d, B = 200, n.c = 100, c.max = 100)
R> plot(vis.d, interactive = FALSE, which = "vip")
R> plot(vis.d, interactive = FALSE, which = "boot", max.circle = 0.25,
+    highlight = "hdl")
R> plot(af.d, interactive = FALSE, best.only = TRUE, 
+    legend.position = "bottomright")
R> plot(af.d, interactive = FALSE, best.only = FALSE)
\end{CodeInput}
\end{CodeChunk}

A striking feature of the variable inclusion plot is the non-monotonic nature of the \code{hdl} line.   As the penalty value increases, and a more parsimonious model is sought, the \code{hdl} variable is selected more frequently while at the same time other variables with similar information are dropped.  Such paths occur when a group of variables contains similar information to another variable.  The \code{hdl} line is a less extreme example of what occurs with $x_8$ in the artificial example (see Figure \ref{plot.vis}).   The path for the age variable lies below the path for the redundant variable, indicating that it does not provide any useful information. The \code{bmi} and \code{ltg} paths are horizontal with a bootstrap probability of 1 for all penalty values indicating that they are very important variables, as are \code{map} and \code{sex}.  From the variable inclusion plot alone, it is not obvious whether \code{tc} or \code{hdl} is the next most important variable.  Some guidance on this issue is provided by the model stability and adaptive fence plots.

In order to determine which circles correspond to which models in the static version of the bootstrap stability plot, we need to consult the print output of the \code{vis} object.

\begin{CodeChunk}
\begin{CodeInput}
R> vis.d
\end{CodeInput}
\begin{CodeOutput}
                     name prob logLikelihood
                      y~1 1.00      -2547.17
                    y~bmi 0.73      -2454.02
                y~bmi+ltg 1.00      -2411.20
            y~bmi+map+ltg 0.69      -2402.61
         y~bmi+map+tc+ltg 0.42      -2397.48
        y~bmi+map+hdl+ltg 0.32      -2397.71
    y~sex+bmi+map+hdl+ltg 0.67      -2390.13
 y~sex+bmi+map+tc+ldl+ltg 0.47      -2387.30
\end{CodeOutput}
\end{CodeChunk}

As in the variable inclusion plots, it is clear that the two most important variables are \code{bmi} and \code{ltg}, and the third most important variable is  \code{map}.  In models of size four (including the intercept), the model with \code{bmi}, \code{ltg} and \code{map} was selected in 69\% of bootstrap resamples.  There is no clear dominant model in models of size five, with \code{tc} and \code{hdl} both competing to be included.  In models of size six, the combination of \code{sex} and \code{hdl} with the core variables \code{bmi}, \code{map} and \code{ltg}, is the most stable option; it is selected in 67\% of bootstrap resamples.  As the size of the model space in dimension six is much larger than the size of the model space for dimension four, it could be suggested that the 0.67 empirical probability for the \{\code{bmi}, \code{map}, \code{ltg}, \code{sex}, \code{hdl}\} model is a stronger result than the 0.69 result for the \{\code{bmi}, \code{ltg}, \code{map}\} model. 

The adaptive fence plots in the bottom row of Figure \ref{fig:diabetesmain} show a clear peak for the model with just \code{bmi} and \code{ltg}. There are two larger models that also occupy regions of stability, albeit with much lower peaks.  These are \{\code{bmi}, \code{map}, \code{ltg}\} and \{\code{bmi}, \code{map}, \code{ltg}, \code{sex}, \code{hdl}\} which also showed up as dominant models in the model stability plots.  Contrasting  \code{best.only = TRUE} in the lower left panel with  \code{best.only = FALSE} in the lower right panel, we can see that the peaks tend to be more clearly distinguished, though the regions of stability remain largely unchanged.

Stepwise approaches using a forward search or backward search with the AIC or BIC all yield a model with \{\code{bmi}, \code{map}, \code{ltg}, \code{sex}, \code{ldl}, \code{tc}\}.  This model was selected 47\% of the time in models of size 7.  The agreement between the stepwise methods may be comforting for the researcher, but it does not aid a discussion about what other models may be worth exploring.

An interactive version of the plots in Figure \ref{fig:diabetesmain} is available at garthtarr.com/apps/mplot.

\begin{figure}[p]
\centering
\includegraphics[width=0.45\textwidth]{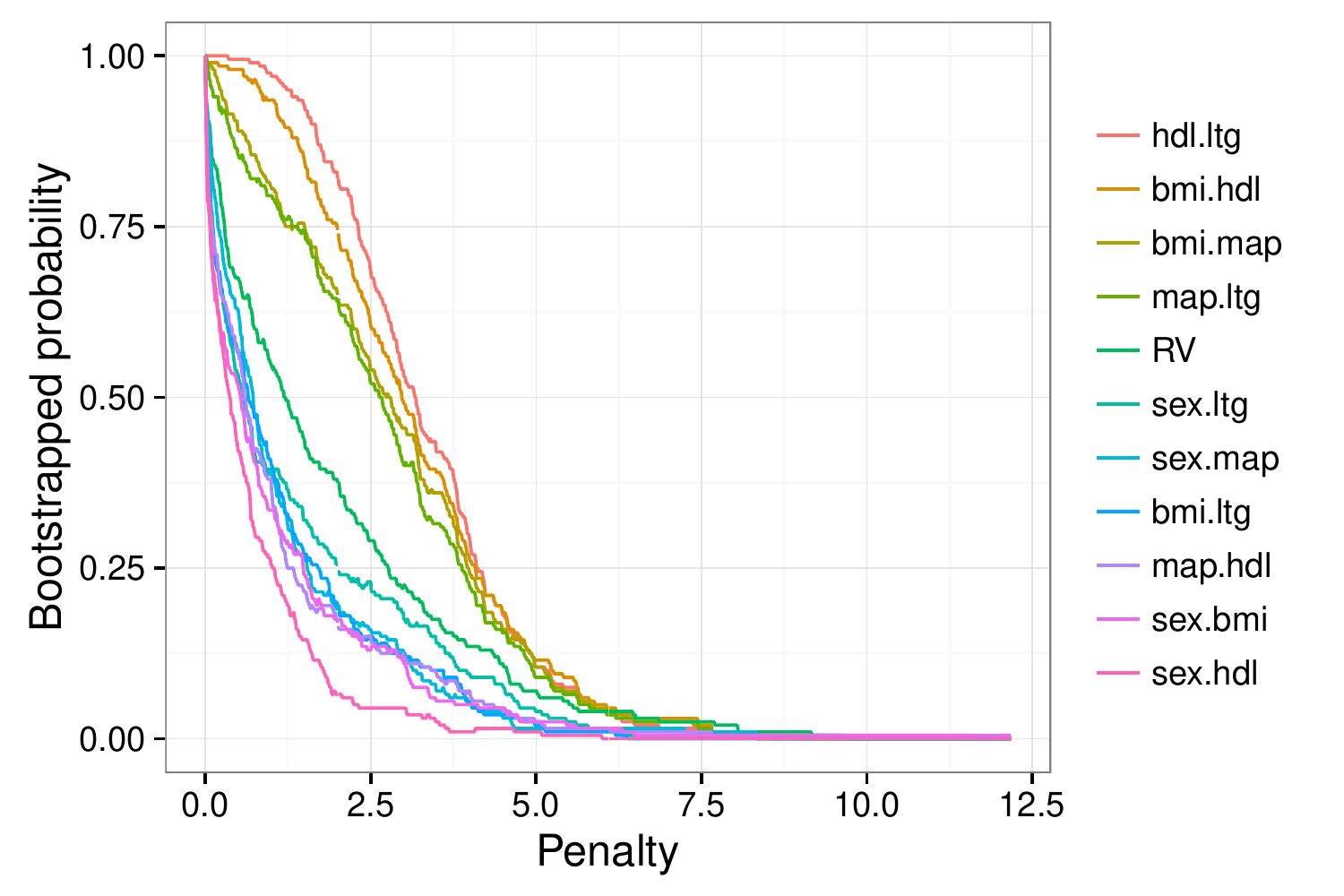}\includegraphics[width=0.45\textwidth]{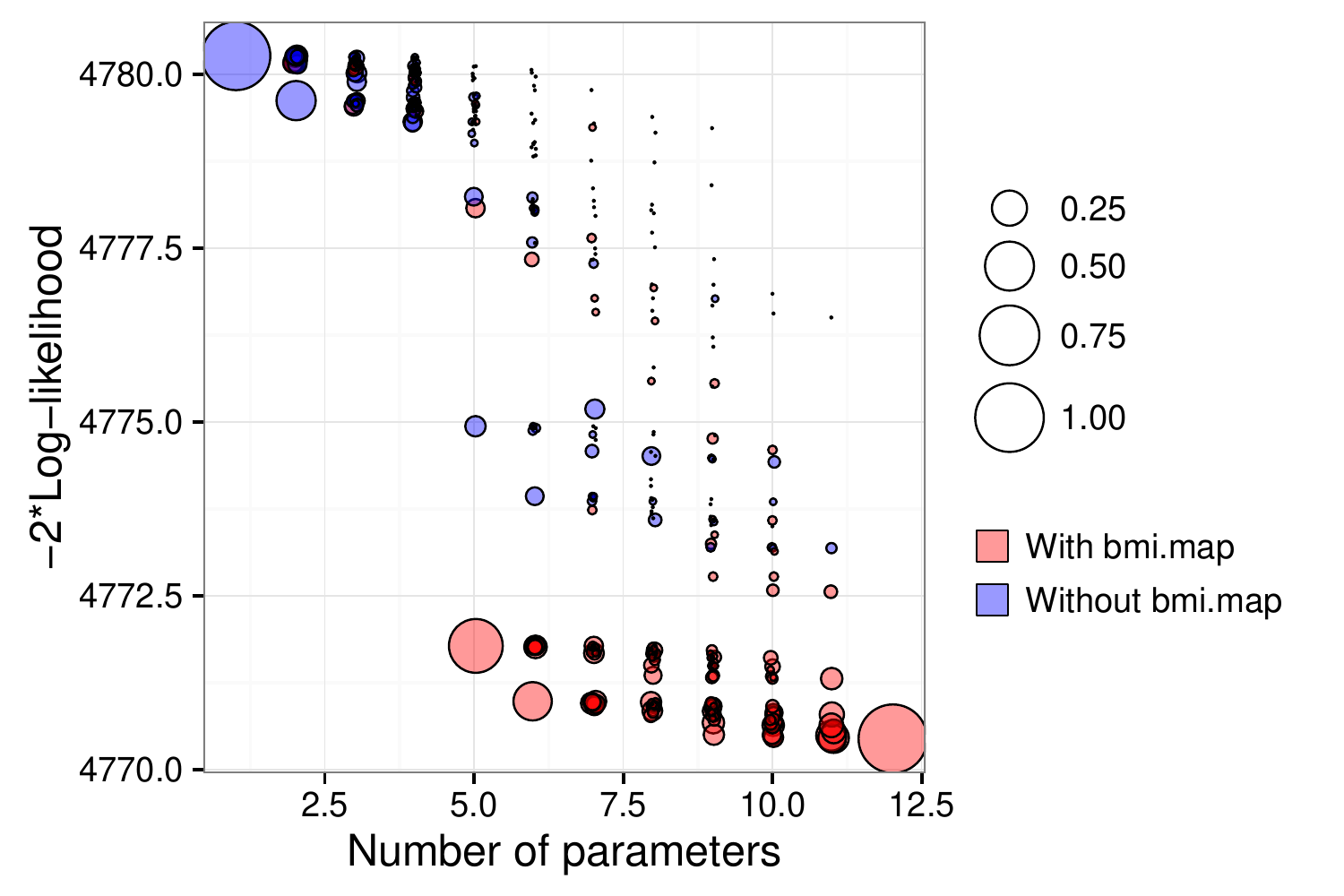}
\includegraphics[width=0.45\textwidth]{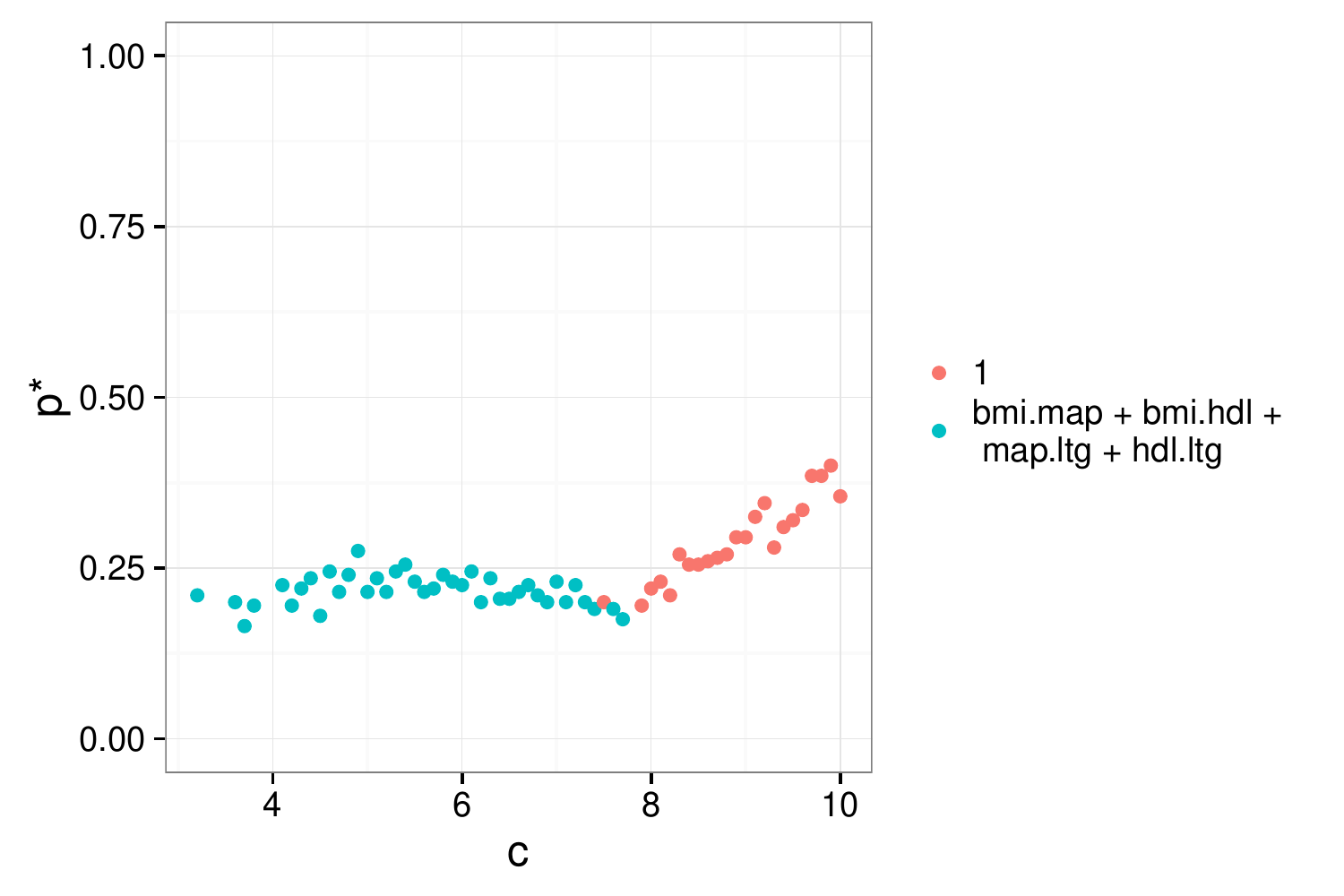}\includegraphics[width=0.45\textwidth]{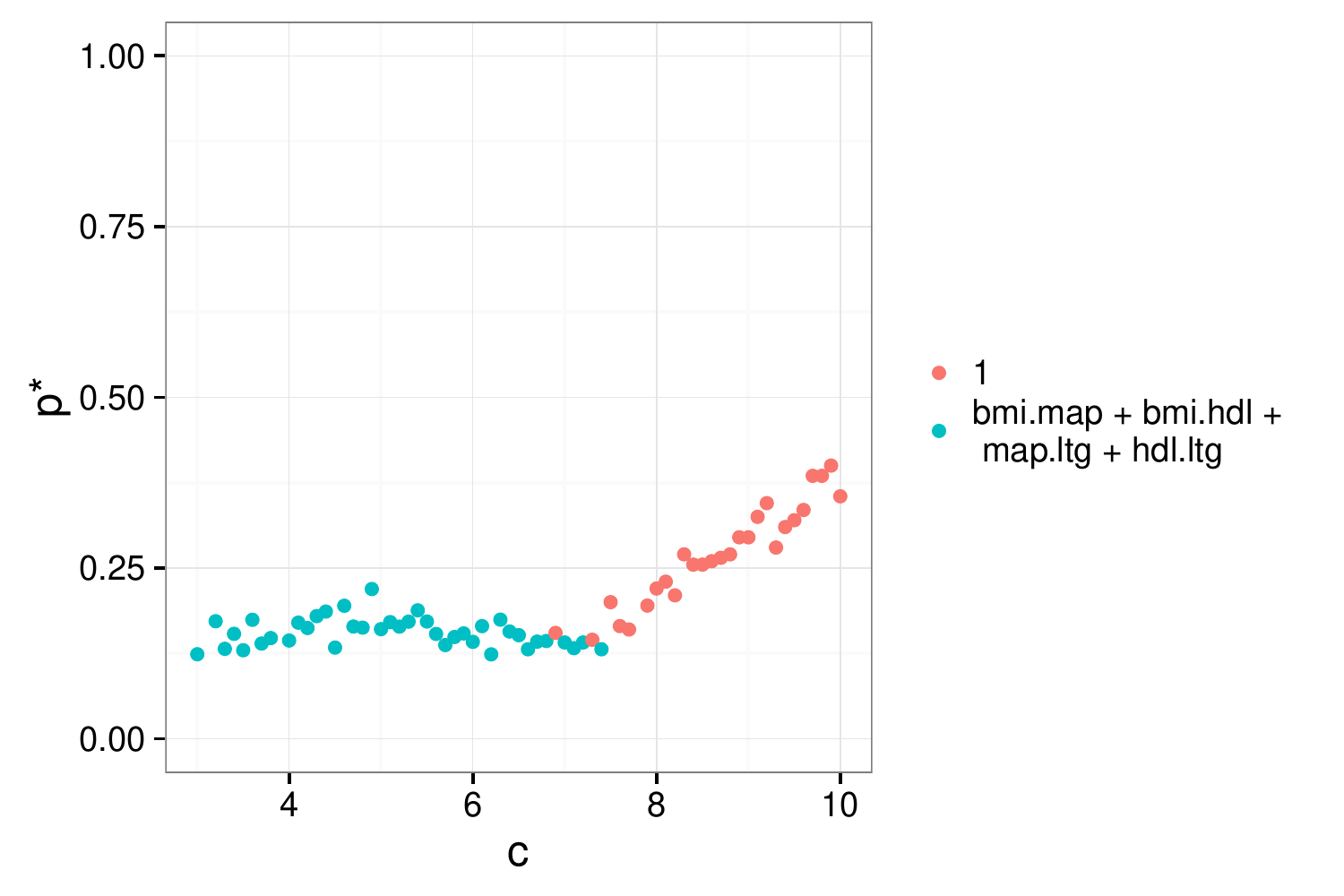}
\caption{Diabetes interactions terms example.}
\label{fig:diabetesint}
\end{figure}

To incorporate interaction terms, we suggest selecting the main effects first, then regressing the relevant interaction terms on the residuals from the main effects model.  This approach ensures that the main effects are always taken into account. In this example, we estimate the dominant model of dimension six and obtain the fitted residuals.  The interaction terms are then regressed on the fitted residuals. 

\begin{CodeChunk}
\begin{CodeInput}
R> lm.d.main = lm(y ~ sex + bmi + map + hdl + ltg, data = diabetes)
R> summary(lm.d.main)
R> db.main = diabetes[, c("sex", "bmi", "map", "hdl", "ltg")]
R> db.main$y = lm.d.main$residuals
R> lm.d.int = lm(y ~ .*. - sex - bmi - map - hdl - ltg, data = db.main)
R> vis.d.int = vis(lm.d.int, B = 200)
R> af.d.int = af(lm.d.int, B = 200, n.c = 100)
R> vis.d.int
\end{CodeInput}
\begin{CodeOutput}
                              name prob logLikelihood
                               y~1 1.00      -2390.13
 y~bmi.map+bmi.hdl+map.ltg+hdl.ltg 0.56      -2385.89
\end{CodeOutput}
\end{CodeChunk}

The result can be found in Figure \ref{fig:diabetesint}. The variable inclusion plots suggest that the most important interaction terms are \code{hdl.ltg}, \code{bmi.hdl}, \code{map.ltg} and \code{bmi.map}.  The model stability plot suggests that there are no dominant models of size 2, 3 or 4.  Furthermore there are no models of size 2, 3 or 4 that make large improvements in description loss. There is a dominant model of dimension 5 that is selected in 56\% of bootstrap resamples.  The variables selected in the dominant model are \{\code{bmi.map}, \code{bmi.hdl}, \code{map.ltg}, \code{hdl.ltg}\}, which can be found in the print output above.  Furthermore, this model does make a reasonable improvement in description loss, almost in line with the full model.  This finding is reinforced in the adaptive fence plots where there are only two regions of stability, one for the null model and another for the \{\code{bmi.map}, \code{bmi.hdl}, \code{map.ltg}, \code{hdl.ltg}\} model. In this instance, the difference between \code{best.only = TRUE} and \code{best.only = FALSE} is minor.

Hence, as a final model for the diabetes example we suggest including the main effects \{\code{bmi}, \code{map}, \code{ltg}, \code{sex}, \code{hdl}\} and the interaction effects \{\code{bmi.map}, \code{bmi.hdl}, \code{map.ltg}, \code{hdl.ltg}\}.  Further investigation can also be useful.  For example, we could use cross validation to compare the model with interaction effects, the model with just main effects and other simpler models that were identified as having peaks in the adaptive fence.  Researchers should also incorporate their specialist knowledge of the predictors and evaluate whether or not the estimated model is sensible from a scientific perspective.

\subsection{Birth weight example}\label{sec:bw}

The second example is the \code{birthwt} dataset from the \pkg{MASS} package which has data on 189 births at the Baystate Medical Centre, Springfield, Massachusetts during 1986 \citep{Venables:2002}.  The main variable of  interest is low birth weight, a binary response variable \code{low} \citep{Hosmer:1989book}.  We have taken the same approach to modelling the full model as in \citet[pp.\ 194--197]{Venables:2002}, where \code{ptl} is reduced to a binary indicator of past history and \code{ftv} is reduced to a factor with three levels.

\begin{CodeChunk}
\begin{CodeInput}
R> require(MASS)
R> bwt <- with(birthwt, {
+    race <- factor(race, labels = c("white", "black", "other"))
+    ptd <- factor(ptl > 0)
+    ftv <- factor(ftv)
+    levels(ftv)[-(1:2)] <- "2+"
+    data.frame(low = factor(low), age, lwt, race, smoke = (smoke > 0),
+      ptd, ht = (ht > 0), ui = (ui > 0), ftv)
+  })
R> options(contrasts = c("contr.treatment", "contr.poly"))
R> bw.glm <- glm(low ~ ., family = binomial, data = bwt)
R> round(summary(bw.glm)$coef, 2)
\end{CodeInput}
\begin{CodeOutput}
            Estimate Std. Error z value Pr(>|z|)
(Intercept)     0.82       1.24    0.66     0.51
age            -0.04       0.04   -0.96     0.34
lwt            -0.02       0.01   -2.21     0.03
raceblack       1.19       0.54    2.22     0.03
raceother       0.74       0.46    1.60     0.11
smokeTRUE       0.76       0.43    1.78     0.08
ptdTRUE         1.34       0.48    2.80     0.01
htTRUE          1.91       0.72    2.65     0.01
uiTRUE          0.68       0.46    1.46     0.14
ftv1           -0.44       0.48   -0.91     0.36
ftv2+           0.18       0.46    0.39     0.69
\end{CodeOutput}
\end{CodeChunk}

\begin{figure}
\centering
\includegraphics[width=0.9\textwidth]{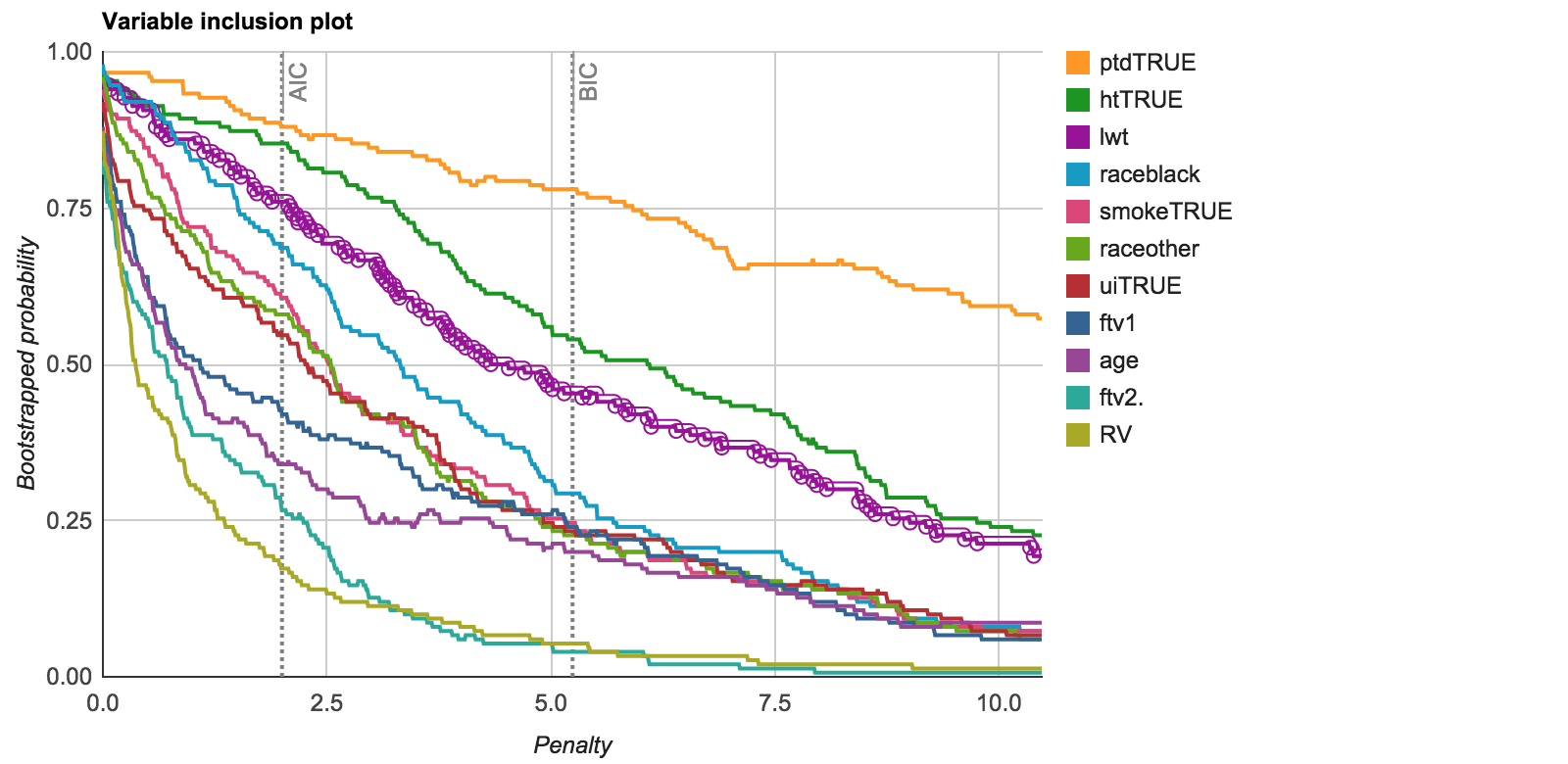}
\includegraphics[width=0.9\textwidth]{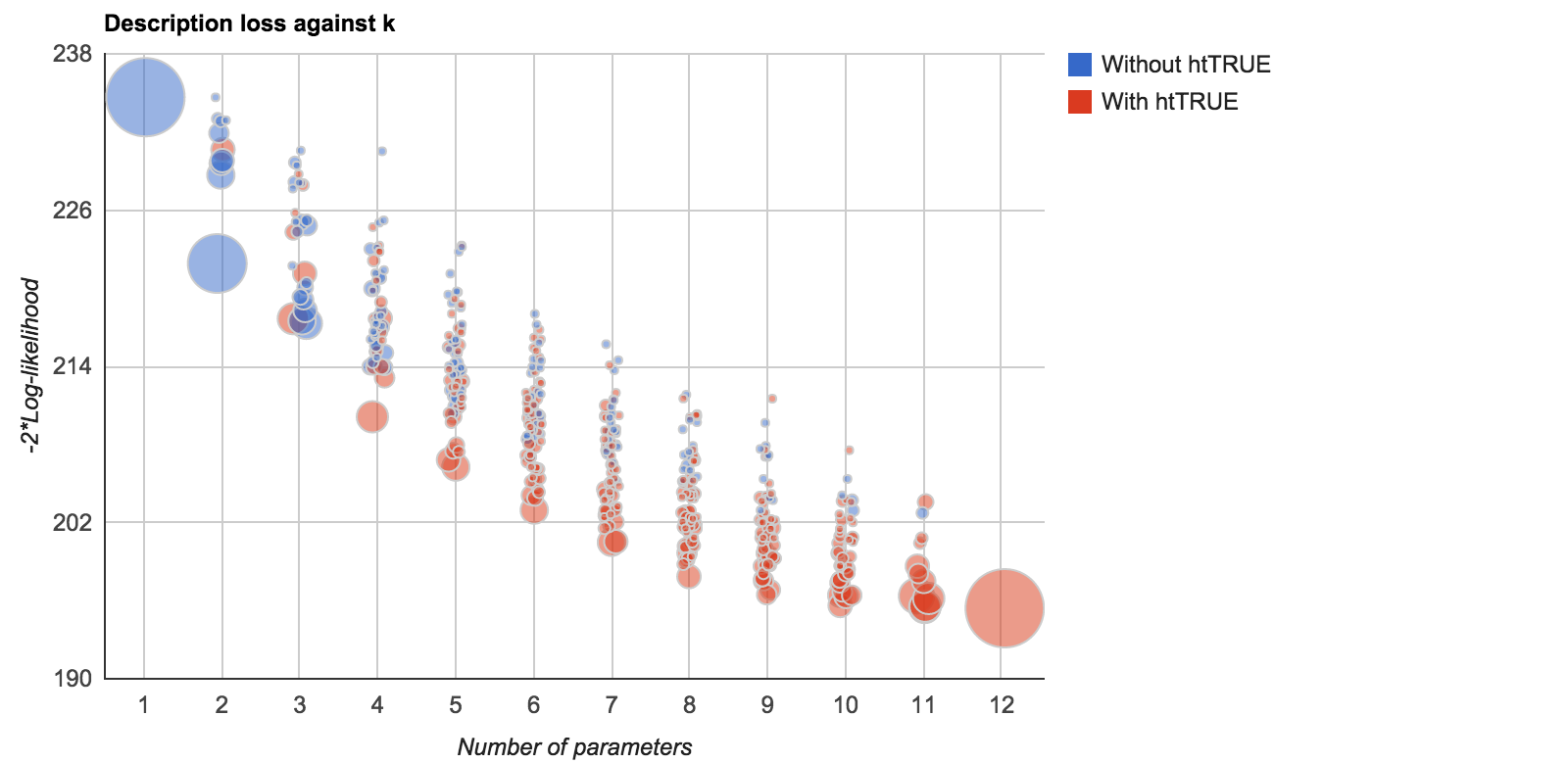}
\includegraphics[width=0.9\textwidth]{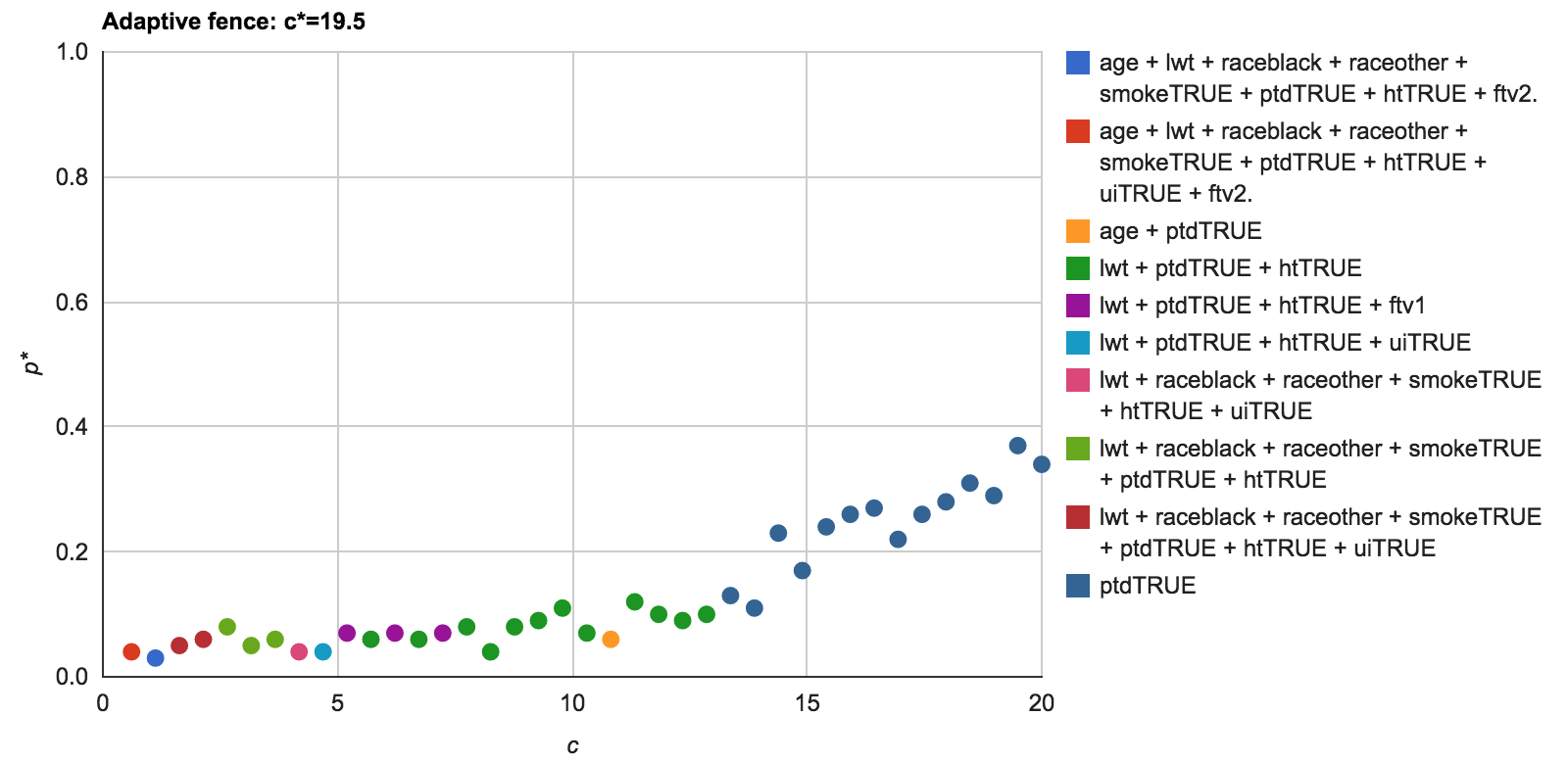}
\caption{Birth weight example.}
\label{fig:birthwt}
\end{figure}

The \code{vis} and \code{af} objects are generated using the fitted full model object as an argument to the \code{vis()} and \code{af()} functions.  The results are shown in Figure \ref{fig:birthwt}, where screenshots have been taken of the interactive plots because they display the larger set of variables more clearly than the static plot methods.
\begin{CodeChunk}
\begin{CodeInput}
R> af.bw = af(bw.glm, B = 150, c.max = 20, n.c = 40)
R> vis.bw = vis(bw.glm, B = 150)
R> plot(vis.bw, which = "vip")
R> plot(vis.bw, which = "boot", highlight = "htTRUE")
R> plot(af.bw)
R> print(vis.bw, min.prob = 0.15)
\end{CodeInput}
\begin{CodeOutput}
                  name prob logLikelihood
                 low~1 1.00       -117.34
           low~ptdTRUE 0.53       -110.95
       low~age+ptdTRUE 0.15       -108.65
low~lwt+ptdTRUE+htTRUE 0.16       -105.06
...
\end{CodeOutput}
\end{CodeChunk}

In this example, it is far less clear which is the best model, or if indeed a ``best model'' exists.  All the curves in the variable inclusion plot lie above the redundant variable curve, with \code{ftv2+} the least important variable.  It is possible to infer an ordering of variable importance from the variable inclusion plots, but there is no clear cutoff as to which variables should be included and which should be excluded.  This is also clear in the model stability plots, where apart from the bivariate regression with \code{ptd}, there are no obviously dominant models. 

In the adaptive fence plot, the only model more complex than a single covariate regression model that shows up with some regularity is the model with \code{lwt}, \code{ptd} and \code{ht}, though at such low levels, it is just barely a region of stability.  This model also stands out slightly in the model stability plot, where it is selected in 16\% of bootstrap resamples and has a slightly lower description loss than other models of the same dimension.  It is worth recalling that the bootstrap resamples generated for the adaptive fence are separate from those generated for the model stability plots.  Indeed the adaptive fence procedure relies on a parametric bootstrap, whereas the model stability plots rely on an exponential weighted bootstrap.  Thus, to find some agreement between these methods is reassuring.

Stepwise approaches using AIC or BIC yield conflicting models, depending on whether the search starts with the full model or the null model.  As expected, the BIC stepwise approach returns smaller models than AIC, selecting the single covariate logistic regression, \code{low ~ ptd}, in the forward direction and the larger model, \code{low ~ lwt + ptd + ht} when stepping backwards from the full model.  Forward selection from the null model with the AIC yielded \code{low ~ ptd + age + ht + lwt + ui} whereas backward selection the slightly larger model, \code{low ~ lwt + race + smoke + ptd + ht + ui}.   Some of these models appear as features in the model stability plots.  Most notably the dominant single covariate logistic regression and the model with \code{lwt}, \code{ptd} and \code{ht} identified as a possible region of stability in the adaptive fence plot.  The larger models identified by the AIC are reflective of the variable importance plot in that they show there may still be important information contained in a number of other variables not identified by the BIC approach.

\citet{Calcagno:2010} also consider this data set, but they allow for the possibility of interaction terms.  Using their approach, they identify ``two'' best models
\begin{CodeChunk}
\begin{CodeOutput}
low ~ smoke + ptd + ht + ui + ftv + age + lwt + ui:smoke + ftv:age
low ~ smoke + ptd + ht + ui + ftv + age + lwt + ui:smoke + ui:ht + ftv:age
\end{CodeOutput}
\end{CodeChunk}
As a general rule, we would warn against the \code{.*.} approach, where all possible interaction terms are considered, as it does not consider whether or not the interaction terms actually make practical sense.  \citet{Calcagno:2010}  conclude that ``Having two best models and not one is an extreme case where taking model selection uncertainty into account rather than looking for a single best model is certainly recommended!''  The issue here is that the software did not highlight that these models are  identical as the \code{ui:ht} interaction variable is simply a vector of ones, and as such, is ignored by the GLM fitting routine.

As computation time can be an issue for GLMs, it is useful to approximate the results using weighted least squares \citep{Hosmer:1989}.  In practice this can be done by fitting the logistic regression and extracting the estimated logistic probabilities, $\hat{\pi}_{i}$.  A new dependent variable is then constructed,
$$z_{i} = \log\left(\frac{\hat{\pi}_{i}}{1-\hat{\pi}_{i}}\right)  + \frac{y_{i}-\hat{\pi}_{i}}{\hat{\pi}_{i}(1-\hat{\pi}_{i})},$$ 
along with observation weights $v_{i}=\hat{\pi}_{i}(1-\hat{\pi}_{i})$. For any submodel $\alpha$ this approach produces the approximate coefficient estimates of \citet{Lawless:1978} and enables us to use the \pkg{leaps} package to perform the computations for best subsets logistic regression as follows.

\begin{CodeChunk}
\begin{CodeInput}
R> pihat = bw.glm$fitted.values
R> r = bw.glm$residuals 
R> z = log(pihat/(1 - pihat)) + r
R> v = pihat*(1 - pihat)
R> nbwt = bwt
R> nbwt$z = z
R> nbwt$low = NULL
R> bw.lm = lm(z ~ ., data = nbwt, weights = v)
R> bw.lm.vis = vis(bw.lm, B = 150)
R> bw.lm.af = af(bw.lm, B = 150 c.max = 20, n.c = 40)
R> plot(bw.lm.vis, which = "vip")
R> plot(bw.lm.vis, which = "boot", highlight = "htTRUE")
R> plot(bw.lm.af)
\end{CodeInput}
\end{CodeChunk}

\begin{figure}
\centering
\includegraphics[width=0.9\textwidth]{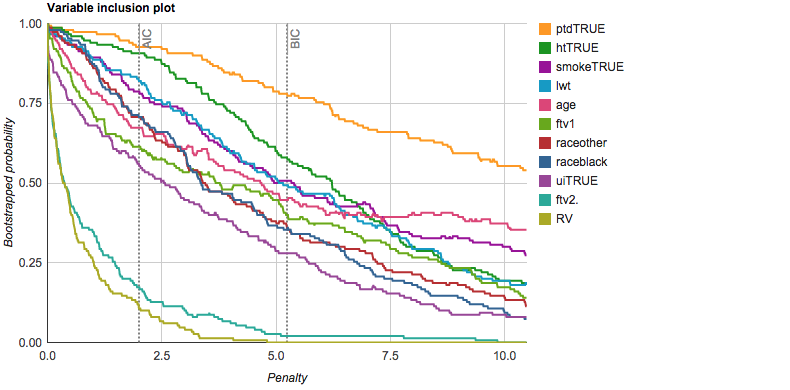}
\includegraphics[width=0.9\textwidth]{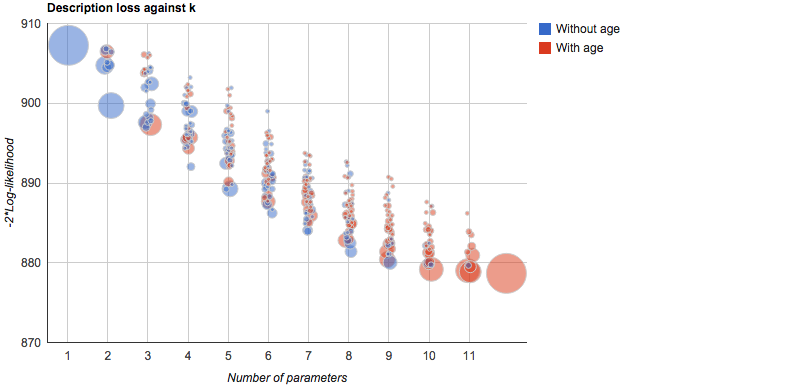}
\includegraphics[width=0.9\textwidth]{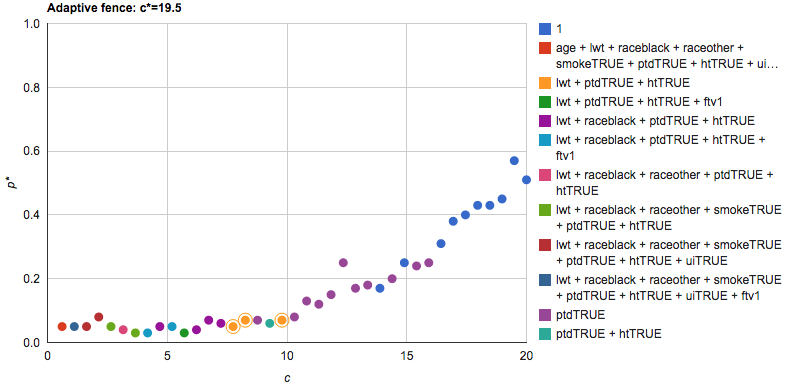}
\caption{Birth weight example with linear model approximation.}
\label{fig:bwtapprox}
\end{figure}

The coefficients from \code{bw.lm}  are identical to \code{bw.glm}.  This approximation provides similar results, shown in Figure \ref{fig:bwtapprox}, in a fraction of the time.

\section{Conclusion}\label{sec:conclusion}

In the rejoinder to their least angle regression paper, \citet{Efron:2004} comment,
\begin{quote}
``In actual practice, or at least in good actual practice, there is a cycle of activity between the investigator, the statistician and the computer \ldots\  The statistician examines the output critically, as did several of our commentators, discussing the results with the investigator, who may at this point suggest adding or removing explanatory variables, and so on, and so on.''
\end{quote}

We hope the suite of methods available in the \pkg{mplot} package adds valuable information to this cycle of activity between researchers and statisticians.  In particular, providing statisticians and researchers alike with a deeper understanding of the relative importance of different models and the variables contained therein.

In the artificial example, we demonstrated a situation where giving the researcher more information in a graphical presentation can lead to choosing the ``correct'' model when standard stepwise procedures would have failed.  

The diabetes data set suggested the existence of a number of different dominant models at various model sizes which could then be investigated further, for example, statistically using cross validation to determine predictive ability, or in discussion with researchers to see which makes the most practical sense.  In contrast, there are no clear models suggested for the birth weight example.  The adaptive fence has no peaks, nor is there a clearly dominant model in the model stability plot even though all but one variable are more informative than the added redundant variable in the variable inclusion plot.

While the core of the \pkg{mplot} package is built around exhaustive searches, this becomes computationally infeasible as the number of variables grows.  We have implemented similar visualisations to model stability plots and variable inclusion plots for \pkg{glmnet} which  brings the concept of model stability to much larger model sizes, though it will no longer be based around exhaustive searches.

The graphs provided by the \pkg{mplot} package are a major contribution.  A large amount of information is generated by the various methods and the best way to interpret that information is through effective visualisations.  For example, as was be shown in Section \ref{sec:diabetes}, the path a variable takes through the variable inclusion plot is often more important than the average inclusion probability over the range of penalty values considered.  It can also be instructive to observe when there are no peaks in the adaptive fence plot as this indicates that the variability of the log-likelihood is limited and no single model stands apart from the others.  Such a relatively flat likelihood over various models would also be seen in the model stability plot where there was no dominant model over the range of model sizes considered.

Although interpretation of the model selection plots provided here is something of an ``art'', this is not something to shy away from.  We accept and train young statisticians to interpret qq-plots and residual plots.  There is a wealth of information in our plots, particularly the interactive versions enhanced with the shiny interface, that can better inform a researchers' model selection choice.

\section*{Acknowledgments}

This research was undertaken with the assistance of resources from the National Computational Infrastructure (NCI), which is supported by the Australian Government.  Samuel Mueller and Alan Welsh were supported by the Australian Research Council (DP140101259).   We also gratefully acknowledge two anonymous referees for their helpful comments and suggestions for the paper and package.

\bibliography{jss3} 

\begin{thebibliography}{38}
\newcommand{\enquote}[1]{``#1''}
\providecommand{\natexlab}[1]{#1}
\providecommand{\url}[1]{\texttt{#1}}
\providecommand{\urlprefix}{URL }
\expandafter\ifx\csname urlstyle\endcsname\relax
  \providecommand{\doi}[1]{doi:\discretionary{}{}{}#1}\else
  \providecommand{\doi}{doi:\discretionary{}{}{}\begingroup
  \urlstyle{rm}\Url}\fi
\providecommand{\eprint}[2][]{\url{#2}}

\bibitem[{Calcagno and de~Mazancourt(2010)}]{Calcagno:2010}
Calcagno V, de~Mazancourt C (2010).
\newblock \enquote{\pkg{glmulti}: An \proglang{R} Package for Easy Automated
  Model Selection with (Generalized) Linear Models.}
\newblock \emph{Journal of Statistical Software}, \textbf{34}(12), 1--29.
\newblock \urlprefix\url{http://www.jstatsoft.org/v34/i12}.

\bibitem[{Chang(2015)}]{Chang:2015b}
Chang W (2015).
\newblock \emph{\pkg{shinydashboard}: Create Dashboards with \pkg{shiny}}.
\newblock \proglang{R} package version 0.5.0,
  \urlprefix\url{http://CRAN.R-project.org/package=shinydashboard}.

\bibitem[{Chang \emph{et~al.}(2015)Chang, Cheng, Allaire, Xie, and
  McPherson}]{Chang:2015a}
Chang W, Cheng J, Allaire J, Xie Y, McPherson J (2015).
\newblock \emph{\pkg{shiny}: Web Application Framework for \proglang{R}}.
\newblock \proglang{R} package version 0.12.1.9000,
  \urlprefix\url{http://shiny.rstudio.com}.

\bibitem[{Efron \emph{et~al.}(2004)Efron, Hastie, Johnstone, and
  Tibshirani}]{Efron:2004}
Efron B, Hastie T, Johnstone I, Tibshirani R (2004).
\newblock \enquote{Least Angle Regression.}
\newblock \emph{The Annals of Statistics}, \textbf{32}(2), 407--451.
\newblock \doi{10.1214/009053604000000067}.

\bibitem[{Friedman \emph{et~al.}(2010)Friedman, Hastie, and
  Tibshirani}]{Friedman:2010}
Friedman JH, Hastie T, Tibshirani R (2010).
\newblock \enquote{Regularization Paths for Generalized Linear Models via
  Coordinate Descent.}
\newblock \emph{Journal of Statistical Software}, \textbf{33}(1), 1--22.
\newblock \urlprefix\url{http://www.jstatsoft.org/v33/i01}.

\bibitem[{Gabry(2015)}]{Gabry:2015}
Gabry J (2015).
\newblock \emph{\pkg{shinystan}: Interactive Visual and Numerical Diagnostics
  and Posterior Analysis for for Bayesian Models}.
\newblock \proglang{R} package version 2.0.0,
  \urlprefix\url{http://CRAN.R-project.org/package=shinystan}.

\bibitem[{Gesmann and de~Castillo(2011)}]{Gesmann:2011}
Gesmann M, de~Castillo D (2011).
\newblock \enquote{Using the {Google} Visualisation {API} with \proglang{R}.}
\newblock \emph{The \proglang{R} Journal}, \textbf{3}(2), 40--44.

\bibitem[{Harrell(2001)}]{Harrell:2001}
Harrell F (2001).
\newblock \emph{Regression Modeling Strategies: With Applications to Linear
  Models, Logistic Regression, and Survival Analysis}.
\newblock Springer-Verlag, New York.

\bibitem[{Hosmer and Lemeshow(1989)}]{Hosmer:1989book}
Hosmer D, Lemeshow S (1989).
\newblock \emph{Applied Logistic Regression}.
\newblock John Wiley \& Sons, New York.

\bibitem[{Hosmer \emph{et~al.}(1989)Hosmer, Jovanovic, and
  Lemeshow}]{Hosmer:1989}
Hosmer DW, Jovanovic B, Lemeshow S (1989).
\newblock \enquote{Best Subsets Logistic Regression.}
\newblock \emph{Biometrics}, \textbf{45}(4), 1265--1270.
\newblock \doi{10.2307/2531779}.

\bibitem[{Jiang(2014)}]{Jiang:2014}
Jiang J (2014).
\newblock \enquote{The Fence Methods.}
\newblock \emph{Advances in Statistics}, \textbf{2014}, 1--14.
\newblock \doi{10.1155/2014/830821}.

\bibitem[{Jiang \emph{et~al.}(2009)Jiang, Nguyen, and Rao}]{Jiang:2009}
Jiang J, Nguyen T, Rao JS (2009).
\newblock \enquote{A Simplified Adaptive Fence Procedure.}
\newblock \emph{Statistics \& Probability Letters}, \textbf{79}(5), 625--629.
\newblock \doi{10.1016/j.spl.2008.10.014}.

\bibitem[{Jiang \emph{et~al.}(2008)Jiang, Rao, Gu, and Nguyen}]{Jiang:2008}
Jiang J, Rao JS, Gu Z, Nguyen T (2008).
\newblock \enquote{Fence Methods for Mixed Model Selection.}
\newblock \emph{The Annals of Statistics}, \textbf{36}(4), 1669--1692.
\newblock \doi{10.1214/07-AOS517}.

\bibitem[{Konishi and Kitagawa(1996)}]{Konishi:1996}
Konishi S, Kitagawa G (1996).
\newblock \enquote{Generalised Information Criteria in Model Selection.}
\newblock \emph{Biometrika}, \textbf{83}(4), 875--890.
\newblock \doi{10.1093/biomet/83.4.875}.

\bibitem[{Lawless and Singhal(1978)}]{Lawless:1978}
Lawless JF, Singhal K (1978).
\newblock \enquote{Efficient Screening of Nonnormal Regression Models.}
\newblock \emph{Biometrics}, \textbf{34}(2), 318--327.
\newblock \doi{10.2307/2530022}.

\bibitem[{Lumley and Miller(2009)}]{Lumley:2009}
Lumley T, Miller A (2009).
\newblock \emph{\pkg{leaps}: Regression Subset Selection}.
\newblock \proglang{R} package version 2.9,
  \urlprefix\url{http://CRAN.R-project.org/package=leaps}.

\bibitem[{Mallows(2000)}]{Mallows:2000}
Mallows CL (2000).
\newblock \enquote{Some Comments on $C_p$.}
\newblock \emph{Technometrics}, \textbf{42}(1), 87--94.
\newblock \doi{10.1080/00401706.2000.10485984}.

\bibitem[{McLeod and Xu(2014)}]{McLeod:2014}
McLeod A, Xu C (2014).
\newblock \emph{\pkg{bestglm}: Best Subset GLM}.
\newblock \proglang{R} package version 0.34,
  \urlprefix\url{http://CRAN.R-project.org/package=bestglm}.

\bibitem[{McMurdie and Holmes(2013)}]{McMurdie:2013}
McMurdie PJ, Holmes S (2013).
\newblock \enquote{\pkg{phyloseq}: An \proglang{R} Package for Reproducible
  Interactive Analysis and Graphics of Microbiome Census Data.}
\newblock \emph{{PLoS ONE}}, \textbf{8}(4), e61217.
\newblock \doi{10.1371/journal.pone.0061217}.

\bibitem[{Meinshausen and B\"{u}hlmann(2010)}]{Meinshausen:2010}
Meinshausen N, B\"{u}hlmann P (2010).
\newblock \enquote{Stability Selection.}
\newblock \emph{Journal of the Royal Statistical Society: Series B (Statistical
  Methodology)}, \textbf{72}(4), 417--473.
\newblock \doi{10.1111/j.1467-9868.2010.00740.x}.

\bibitem[{Miller(2002)}]{Miller:2002}
Miller A (2002).
\newblock \emph{Subset Selection in Regression}.
\newblock CRC Monographs on Statistics \& Applied Probability. Chapman \& Hall,
  Boca Raton.

\bibitem[{M\"{u}ller \emph{et~al.}(2013)M\"{u}ller, Scealy, and
  Welsh}]{Mueller:2013}
M\"{u}ller S, Scealy JL, Welsh AH (2013).
\newblock \enquote{Model Selection in Linear Mixed Models.}
\newblock \emph{Statistical Science}, \textbf{28}(2), 135--167.
\newblock \doi{10.1214/12-STS410}.

\bibitem[{M{\"u}ller and Vial(2009)}]{Mueller:2009b}
M{\"u}ller S, Vial C (2009).
\newblock \enquote{Partially Linear Model Selection by the Bootstrap.}
\newblock \emph{Australian \& New Zealand Journal of Statistics},
  \textbf{51}(2), 183--200.
\newblock \doi{10.1111/j.1467-842X.2009.00540.x}.

\bibitem[{M\"{u}ller and Welsh(2010)}]{Mueller:2010}
M\"{u}ller S, Welsh A (2010).
\newblock \enquote{On Model Selection Curves.}
\newblock \emph{International Statistical Review}, \textbf{78}(2), 240--256.
\newblock \doi{10.1111/j.1751-5823.2010.00108.x}.

\bibitem[{M\"{u}ller and Welsh(2005)}]{Mueller:2005}
M\"{u}ller S, Welsh AH (2005).
\newblock \enquote{Outlier Robust Model Selection in Linear Regression.}
\newblock \emph{Journal of the American Statistical Association},
  \textbf{100}(472), 1297--1310.
\newblock \doi{10.1198/016214505000000529}.

\bibitem[{M\"{u}ller and Welsh(2009)}]{Mueller:2009}
M\"{u}ller S, Welsh AH (2009).
\newblock \enquote{Robust Model Selection in Generalized Linear Models.}
\newblock \emph{Statistica Sinica}, \textbf{19}(3), 1155--1170.

\bibitem[{Murray \emph{et~al.}(2013)Murray, Heritier, and
  M\"{u}ller}]{Murray:2013}
Murray K, Heritier S, M\"{u}ller S (2013).
\newblock \enquote{Graphical Tools for Model Selection in Generalized Linear
  Models.}
\newblock \emph{Statistics in Medicine}, \textbf{32}(25), 4438--4451.
\newblock \doi{10.1002/sim.5855}.

\bibitem[{Park \emph{et~al.}(2014)Park, Sakaori, and Konishi}]{Park:2014}
Park H, Sakaori F, Konishi S (2014).
\newblock \enquote{Robust Sparse Regression and Tuning Parameter Selection via
  the Efficient Bootstrap Information Criteria.}
\newblock \emph{Journal of Statistical Computation and Simulation},
  \textbf{84}(7), 1596--1607.
\newblock \doi{10.1080/00949655.2012.755532}.

\bibitem[{{\proglang{R} Core Team}(2015)}]{R-Core-Team:2015aa}
{\proglang{R} Core Team} (2015).
\newblock \emph{\proglang{R}: A Language and Environment for Statistical
  Computing}.
\newblock \proglang{R} Foundation for Statistical Computing, Vienna, Austria.
\newblock \urlprefix\url{https://www.R-project.org/}.

\bibitem[{{Revolution Analytics} and
  Weston(2014{\natexlab{a}})}]{doParallel:2014}
{Revolution Analytics}, Weston S (2014{\natexlab{a}}).
\newblock \emph{\pkg{doParallel}: Foreach Parallel Adaptor for the Parallel
  Package}.
\newblock \proglang{R} package version 1.0.8,
  \urlprefix\url{http://CRAN.R-project.org/package=doParallel}.

\bibitem[{{Revolution Analytics} and Weston(2014{\natexlab{b}})}]{foreach:2014}
{Revolution Analytics}, Weston S (2014{\natexlab{b}}).
\newblock \emph{\pkg{foreach}: Foreach Looping Construct for \proglang{R}}.
\newblock \proglang{R} package version 1.4.2,
  \urlprefix\url{http://CRAN.R-project.org/package=foreach}.

\bibitem[{Shang and Cavanaugh(2008)}]{Shang:2008}
Shang J, Cavanaugh JE (2008).
\newblock \enquote{Bootstrap Variants of the {Akaike} Information Criterion for
  Mixed Model Selection.}
\newblock \emph{Computational Statistics {\&} Data Analysis}, \textbf{52}(4),
  2004--2021.
\newblock \doi{http://dx.doi.org/10.1016/j.csda.2007.06.019}.

\bibitem[{Shao(1996)}]{Shao:1996}
Shao J (1996).
\newblock \enquote{Bootstrap Model Selection.}
\newblock \emph{Journal of the American Statistical Association},
  \textbf{91}(434), 655--665.
\newblock \doi{10.2307/2291661}.

\bibitem[{Shen \emph{et~al.}(2012)Shen, Pan, and Zhu}]{Shen:2012}
Shen X, Pan W, Zhu Y (2012).
\newblock \enquote{Likelihood-Based Selection and Sharp Parameter Estimation.}
\newblock \emph{Journal of the American Statistical Association},
  \textbf{107}(497), 223--232.
\newblock \doi{10.1080/01621459.2011.645783}.

\bibitem[{Tarr \emph{et~al.}(2016)Tarr, M\"{u}ller, and Welsh}]{Tarr:2015c}
Tarr G, M\"{u}ller S, Welsh A (2016).
\newblock \emph{\pkg{mplot}: Graphical Model Stability and Model Selection
  Procedures}.
\newblock \proglang{R} package version 0.7.9,
  \urlprefix\url{http://CRAN.R-project.org/package=mplot}.

\bibitem[{Tibshirani(1996)}]{Tibshirani:1996}
Tibshirani R (1996).
\newblock \enquote{Regression Shrinkage and Selection via the Lasso.}
\newblock \emph{Journal of the Royal Statistical Society: Series B
  (Methodological)}, pp. 267--288.

\bibitem[{Tibshirani \emph{et~al.}(2004)Tibshirani, Johnstone, Hastie, and
  Efron}]{Tibshirani:2004}
Tibshirani RJ, Johnstone I, Hastie T, Efron B (2004).
\newblock \enquote{Least Angle Regression.}
\newblock \emph{The Annals of Statistics}, \textbf{32}(2), 407--499.
\newblock \doi{10.1214/009053604000000067}.

\bibitem[{Venables and Ripley(2002)}]{Venables:2002}
Venables WN, Ripley BD (2002).
\newblock \emph{Modern Applied Statistics with \proglang{S}}.
\newblock Fourth edition. Springer-Verlag, New York.

\end{thebibliography}

\end{document}